\documentclass[10pt,a4paper,sort&compres]{article}
\usepackage[utf8]{inputenc}
\usepackage[T1]{fontenc}
\usepackage[a4paper,margin=2.5cm]{geometry}
\usepackage{physics}
\usepackage{amsmath,amsthm,amssymb,amsfonts,bm}
\usepackage{graphicx}
\usepackage[font=small,labelfont=bf]{caption}
\usepackage[colorlinks=true,linkcolor=blue,citecolor=blue,urlcolor=blue]{hyperref}
\usepackage{color}
\usepackage{authblk}
\usepackage[backend=biber, sorting=none]{biblatex}
\bibliography{main}

\DeclareMathAlphabet{\pazocal}{OMS}{zplm}{m}{n}
\newcommand{\Ma}{\mathcal{M}}
\newcommand{\Da}{\mathcal{D}}
\newcommand{\Ba}{\mathcal{B}}
\newcommand{\Ha}{\mathcal{H}}

\newcommand{\Oa}{\mathcal{O}}

\newcommand{\de}{{\rm d}}

\title{\textbf{Guided Waves in Static Curved Spacetimes}}

\author[1,2,3]{\textbf{\large{Javier Blanco-Romero}}}
\affil[1]{Department of Telematic Engineering, Universidad Carlos III de Madrid, Leganés, Madrid, 28911, Spain}
\affil[2]{Department of Optics, Faculty of Physical Sciences, Universidad Complutense de Madrid, Pza. Ciencias 1, 28040 Madrid, Spain (2017-2018)}
\affil[3]{E-mail: fjavib@gmail.com}

\begin{document}

\maketitle
		
\begin{abstract}
This work investigates the propagation of electromagnetic waves in waveguides within static curved spacetimes. We develop a covariant formalism using Hertzian potentials to describe guided electromagnetic modes in spacetimes with metrics that depend on the proper length along the waveguide axis. The Maxwell equations are solved using a Hertzian potential ansatz, resulting in wave equations for TE and TM modes. We analyze the axial and transverse components of the solutions and derive expressions for the cutoff frequency and guide wavelength. The special case of TEM modes is also examined. As an illustrative example, we apply the formalism to radial propagation in Schwarzschild spacetime. This work provides a framework for studying guided electromagnetic waves in curved spacetime geometries, opening up potential applications in precision tests of general relativity and relativistic quantum optics.
\end{abstract}

\date{}

\pagebreak

\tableofcontents

\pagebreak

\section{Introduction}

The study of electromagnetic waves in curved spacetime has been a subject of interest since the early days of general relativity. General relativity has a solid experimental basis, with one of its most frequently measured predictions being time dilation due to the presence of a gravitational field. Some of the most representative experiments in this area include \cite{Pound,Hafele_1,Hafele_2,Chou_clocks_relativity}. Notably, all experiments carried out to date have in common that general relativity effects are observed in classical degrees of freedom \cite{Zych_interferometry,Zych_thesis}.

On the other hand, Newtonian gravity effects have also been observed in quantum systems. The neutron interference experiments, among which the pioneering work of \cite{COW} stands out, and the neutron rebound experiments \cite{Nesvizhevsky_theory,Nesvizhevsky_experiment,Jenke} constitute proof that Newtonian gravity plays the same role in quantum mechanics as other external fields and is capable of inducing interference effects and state discretization. However, to date, there are no experiments that simultaneously manifest both general relativity and quantum effects \cite{Zych_interferometry,Zych_thesis}.

This observational gap, combined with the rapid development of technologies that allow working with quantum states over large distances, has made the intersection of classical gravity and quantum mechanics an active area of research. Recent advancements in long-distance optical communications and quantum information transfer have expanded the possibilities for studying fundamental physics over extended scales. Kilometer-scale optical fibers are now routinely used in telecommunications \cite{agrawal2012fiber}, with submarine cables spanning ocean basins to form the backbone of global internet infrastructure. Experimental successes in long-distance quantum communications \cite{yin2017satellite} have further pushed the boundaries of quantum state manipulation over significant distances. These developments, coupled with increasing precision in quantum experiments \cite{Nicholson_atomic_clock}, have opened new possibilities for both scientific and technological applications in this field (see review \cite{howl2018gravity}).

These vast distances suggest the possibility of observing non-local phenomena predicted by general relativity in such systems. Moreover, these extended optical systems are finding novel applications beyond communications. Recent research has demonstrated that submarine fiber optic cables can be repurposed as distributed sensors for detecting earthquakes and studying ocean dynamics \cite{lindsey2019illuminating}. This innovative use of fiber optic infrastructure hints at the potential for applying electromagnetic wave propagation in new and unexpected ways, including the study of gravitational effects over large scales.

The propagation of electromagnetic waves in curved spacetime has been studied extensively in the context of astrophysical phenomena \cite{MTW_Gravitation, synge1960relativity, chandrasekhar1998mathematical}. However, the application of these principles to guided waves in extended terrestrial systems presents both unique challenges and opportunities. This work explores the implications of curved spacetime on guided electromagnetic waves in extended waveguides, with potential applications ranging from precision measurements to quantum sensing.

The interaction between gravity and electromagnetism in waveguides could lead to observable effects in precision measurements. For example, the Sagnac effect in ring lasers has been used to detect Earth's rotation and other inertial effects \cite{stedman1997ring}. Extending this concept to gravitational effects could open new avenues for testing general relativity and alternative theories of gravity \cite{will2018theory}. Furthermore, understanding how spacetime curvature affects the propagation of quantum states of light could advance the fields of relativistic quantum optics and quantum metrology \cite{Rideout2012_quantum_optics_experiments_satellites}, potentially leading to new quantum sensing technologies and improved tests of fundamental physics.

These considerations invite us to explore experimental scenarios where effects of general relativity and quantum mechanics appear simultaneously, allowing us to address this observational gap in the coming years. One such proposal in the literature is the interference of quantum clocks in the Earth's gravitational field \cite{Zych_interferometry,Zych_thesis}.

This work aims to develop a framework for analyzing guided electromagnetic waves in static curved spacetimes using Hertzian potentials. By extending methods developed for flat spacetime \cite{collin1960field} to curved geometries, we focus on static spacetimes as a first step towards more general scenarios. This approach provides a foundation for future studies of electromagnetic wave propagation in more complex gravitational environments, bridging the gap between theoretical astrophysics and practical terrestrial applications.

The rest of this paper is organized as follows:
In Section \ref{theoretical_background_section}, we provide an overview of the theoretical framework required for our analysis. This includes a review of electromagnetism in curved spacetimes and the theory of Hertzian potentials.
Section \ref{guided_waves_static_st_section} forms the core of our work, focusing on guided waves in static spacetimes. We begin by discussing guided waves in flat spacetime using Hertzian potentials, then introduce our waveguide model and the induced metric. We develop the 3+1 field theory and derive the relevant Maxwell and wave equations. We then apply the Hertzian potential formalism to guided waves in static spacetimes, analyzing TE, TM, and TEM modes. We particularize our results to the Schwarzschild spacetime as an example. The section concludes with a discussion on interferometry and general superposition of waves.
Finally, Section \ref{conclusions_section} presents the conclusions of our study, summarizing our key findings and discussing potential implications and future directions for this research.

\subsection{Conventions}
In this paper, we adopt the following conventions:
\begin{itemize}
\item We use the $(-+++)$ signature for the metric.
\item Greek letters ($\alpha, \beta, \mu, \nu, \ldots$) denote spacetime indices, while Latin letters ($a, b, i, j, k, \ldots$) represent spacelike indices.
\item Four-vectors are denoted by italic letters (e.g., $x, y$), while spatial three-vectors are represented by boldface letters (e.g., $\mathbf{v}, \mathbf{E}$).
\item We employ natural units ($c = 1$) and Lorentz-Heaviside units ($\epsilon_0 = \mu_0 = 1$).
\end{itemize}

\pagebreak

\section{Theoretical Background}\label{theoretical_background_section}

Here, we review some of the theoretical groundwork necessary for understanding electromagnetic waves in curved spacetimes. We begin with an overview of electromagnetism in curved geometries, employing both index notation and differential forms to provide a thorough perspective. We then introduce the powerful concept of Hertzian potentials, presenting both vector and covariant formulations and extending their application to curved spacetimes.

\subsection{Electromagnetism in Curved Spacetimes}

\subsubsection{Coordinate-based Formulation}

To describe electromagnetism in curved spacetimes, we begin with the following definitions:

\begin{equation}\label{Electric_field_definition}
    F_{0i} = - E_i \ ,
\end{equation}

\begin{equation}\label{Magnetic_induction_definition}
    F_{jk} =  \epsilon_{jki}B^i \ ,
\end{equation}

Here, $E_i$ represents the components of the electric field, and $B^i$ the components of the magnetic field. Using these definitions, we can express the Faraday 2-form in Minkowski spacetime. Its components are given by \cite{MTW_Gravitation}:

\begin{equation}\label{Faraday_2_form_components}
\left(F_{\mu \nu} \right) = \left(\begin{array}[hbtp]{cccc}
0&-E_x/c&-E_y/c&-E_z/c\\
E_x/c&0&B_z&-B_y\\
E_y/c&-B_z&0&B_x\\
E_z/c&B_y&-B_x&0
\end{array}\right) \ ,
\end{equation}

The dual of the Faraday tensor, denoted as $\star F^{\mu \nu}$, is defined by $\star F^{\mu \nu} = \frac{1}{2}\epsilon^{\mu\nu\rho\sigma}F_{\rho\sigma}$. Its components are:

\begin{equation}\label{Dual_Faraday_2_form_components}
\left(\star F_{\mu \nu} \right) =  \left(\begin{array}[hbtp]{cccc}
0&B^1&B^2&B^3\\
-B^1&0&E^3 /c&-E^2 /c\\
-B^2&-E^3 /c&0&E^1 /c\\
-B^3&E^2 /c&-E^1 /c&0
\end{array}\right) \ ,
\end{equation}

In curved spacetime, the Maxwell equations can be derived from the following action:

\begin{equation}\label{action_components}
S\left[\mathbf{F}, \mathbf{g} \right] = -\frac{1}{4}\int \sqrt{g} \de ^4 x F_{\mu\nu}F^{\mu\nu} \ .
\end{equation}

Here, $g$ represents the determinant of the metric tensor. From this action, we obtain the Maxwell equations in vacuum \cite{Cabral_electrodynamics}:
 
\begin{equation}
\begin{aligned}
\nabla_{[\alpha}F_{\beta \gamma]} &= 0 \ , \\
\nabla_\mu F^{\mu \nu} &= 0 \ .
\end{aligned}
\label{Maxwell_equations_vacuum}
\end{equation}

where $\nabla_\mu$ denotes the covariant derivative. The covariant derivative of the Faraday tensor is given by:

\begin{equation}
\nabla_\mu F^{\lambda \nu} = \partial_\mu F^{\lambda \nu} + \Gamma^\lambda_{\phantom{\lambda}\alpha\mu}F^{\alpha \nu} + \Gamma^\nu_{\phantom{\nu}\alpha\mu}F^{\lambda \alpha}
\label{covariant_inhomogeneous}
\end{equation}

The homogeneous equations \eqref{Maxwell_equations_vacuum} yield the Faraday law and the magnetic Gauss law. These laws are independent of the metric, assuming the spacetime is torsionless. As a result, the Faraday law retains its familiar form:

\begin{equation}\label{Faraday_law}
\partial_t \mathbf{B} = - \nabla \times \mathbf{E} \ ,
\end{equation}

By the other hand, the magnetic Gauss law reads:

\begin{equation}\label{Magnetic_Gauss_law}
    \nabla \cdot \mathbf{B} = 0 \ .
\end{equation}

The inhomogeneous Maxwell equation \eqref{covariant_inhomogeneous} can be simplified by considering the symmetries of both the Christoffel symbols and the Faraday tensor. The contraction of the lower symmetric index of the Christoffel symbols with the upper antisymmetric index of the Faraday tensor in the covariant derivative vanishes. This allows us to write:

\begin{equation}\label{Maxwell_equations_vacuum_inhomogeneous}
    \nabla_\mu F^{\mu \nu} = \partial_\mu F^{\mu \nu} + \Gamma^\mu_{\phantom{\mu}\alpha\mu}F^{\alpha \nu} \ .
\end{equation}

This formulation of Maxwell's equations in curved spacetime provides the foundation for our subsequent analysis of electromagnetic waves in waveguides within static curved spacetimes.

\subsubsection{Differential Forms Formulation}

An alternative and more geometrical approach to electromagnetism in curved spacetime involves the use of differential forms. In this formulation, the Faraday 2-form can be expressed as (see for example \cite{MTW_Gravitation} or \cite{cohen1974electromagnetic}):

\begin{equation}\label{Faraday_2_form}
\mathbf{F} = \frac{1}{2}F_{\alpha \beta} \de x^\alpha \wedge \de x^\beta \ .
\end{equation}

In Minkowski spacetime, this Faraday 2-form reduces to:

\begin{equation}\label{Faraday_2_form_Minkowski}
\begin{split}
\mathbf{F} = &- E_x \de t \wedge \de x - E_y \de t \wedge \de y - E_z \de t \wedge \de z \\
& + B_x \de y \wedge \de z - B_y \de x \wedge \de z + B_z \de x \wedge \de y \\
= &- E_x \de t \wedge \de x - E_y \de t \wedge \de y - E_z \de t \wedge \de z \\
& - B_x \star \de t \wedge \de x - B_y \star \de t \wedge \de y - B_z \star\de t \wedge \de z \ ,
\end{split}
\end{equation}

which is equivalent to the matrix form presented in equation \eqref{Faraday_2_form_components}.

In this formalism, the source-free Maxwell equations take the elegant form:

\begin{equation}\label{Maxwell_equations_coordinate_independent}
\begin{split}
\de \mathbf{F} &= 0  \ , \\
\delta \mathbf{F} &= 0 \ .
\end{split}
\end{equation}

Here, $\de$ represents the exterior derivative and $\delta$ is its adjoint.

The action of the electromagnetic field on a (pseudo-) Riemannian manifold $\Ma$ can be written using differential forms as:

\begin{equation}\label{EM_field_action_with_forms}
S = - \int_\Ma \left(\frac{1}{2} \mathbf{F} \wedge \star \mathbf{F} + \mathbf{A} \wedge \star \mathbf{J} \right) \ ,
\end{equation}

where $\mathbf{J}$ is the current 1-form and $\star$ denotes the Hodge star operator.

To demonstrate how this action relates to the previously introduced form \eqref{action_components}, we expand the term $\frac{1}{2} F \wedge *F$ in components:

\begin{equation}
\begin{split}
\frac{1}{2} F \wedge *F &= \frac{1}{2} \left( \frac12 F_{\mu\nu} dx^\mu \wedge dx^\nu \right) \wedge \left( \frac{1}{2} F^{\alpha \beta} \frac{1}{4!} \epsilon_{\alpha \beta \rho \sigma} dx^\rho \wedge dx^\sigma \right) \\
& = \frac{1}{8} \cdot \frac{1}{4!} F_{\mu\nu} F^{\alpha \beta} \epsilon_{\alpha \beta \rho \sigma} dx^{\mu} \wedge dx^{\nu} \wedge dx^{\rho} \wedge dx^{\sigma} \\
& = \frac{1}{4} \cdot \frac{1}{4!} F_{\alpha \beta} F^{\alpha \beta} \epsilon_{\mu \nu \rho \sigma} dx^{\mu} \wedge dx^{\nu} \wedge dx^{\rho} \wedge dx^{\sigma} \\
& = \frac{1}{4} F_{\alpha \beta} F^{\alpha \beta} \omega \ ,
\end{split}
\end{equation}

where $\omega = d^4x$ is the volume form.

To derive the Maxwell equations from the action \eqref{EM_field_action_with_forms}, we perform a variation of the action induced by variations of the fields $\mathbf{F}$ and $\mathbf{A}$:

\begin{equation}
\delta S = -\frac{1}{2} \int \left( \delta F \wedge \star F + F \wedge \star \delta F \right) + \int \delta A \wedge \star J \ ,
\end{equation}

Assuming a fixed metric, the variation commutes with the Hodge dual. Using the symmetry of the inner product (see \cite{hehl2012foundations}) we have that $\left(F, \delta F \right) = \left(\delta F, F \right)$, so we can simplify this to:
 
\begin{equation}
\delta S = -\int \left( \delta F \wedge \star F  \right) + \int \delta A \wedge \star J \ .
\end{equation}

Expressing this in terms of the four-potential:

\begin{equation}
\delta S = -\int d\delta A \wedge \star dA + \int \delta A \wedge  \star J \ ,
\end{equation}

which, after integration by parts, becomes:

\begin{equation}\label{variation_EM_action}
\delta S = -\int \delta A \wedge d \star dA + \int \delta A \wedge \star J \ .
\end{equation}

The principle of least action requires:

\begin{equation}
\delta S = 0 \quad \forall \ \delta A \ ,
\end{equation}

This, combined with equation \eqref{variation_EM_action}, implies:

\begin{equation}
d \star dA = \star J \ ,
\end{equation}

which are the inhomogeneous Maxwell equations in differential form notation.

\subsubsection{Electromagnetism in Material Media}

To describe electromagnetic phenomena in material media, we adopt the approach outlined in [pg.369, E.3]\cite{hehl2012foundations}. This method involves separating charge and current densities into bounded and external components, providing understanding of electromagnetic interactions within materials.

We begin by splitting the charge and current densities:

\begin{equation}\label{source_splitting}
\begin{split}
\rho &= \rho^{mat} + \rho^{ext} \\
j &= j^{mat} + j^{ext} \ .
\end{split}
\end{equation}

Here, the superscripts 'mat' and 'ext' denote the material (bounded) and external components, respectively.

Assuming a conservation law for the bounded source:

\begin{equation}
\de J ^{mat} = \de j^{mat} + \partial_t \rho^{mat} = 0 \ ,
\end{equation}

we can define a potential that generates the bound sources, analogous to the treatment of free sources:

\begin{equation}\label{bounded_sources_potential}
\de G^{mat} = J^{mat} \ .
\end{equation}

In a (3+1) formalism, this leads to the introduction of the \textbf{polarization} 2-form and the \textbf{magnetization} 1-form:

\begin{equation}\label{polarization_magnetization}
\begin{split}
\Da^{mat} &\equiv - P \\
\Ha^{mat} &\equiv M \ ,
\end{split}
\end{equation}

such that:

\begin{equation}
G^{mat} = \de \sigma \wedge \Ha^{mat} + \Da^{mat} \ .
\end{equation}

These forms satisfy the following relations:

\begin{equation}\label{polarization_magnetization_Maxwell_eqs}
\begin{split}
- \de P &= \rho^{mat} \\
\de M + \partial_t P &= j^{mat} \ .
\end{split}
\end{equation}

Note that these definitions become unique under the assumption that $\Da^{mat} = 0$ when $E=0$ and $\Ha = 0$ when $B= 0$.

Exploiting the linearity of Maxwell's equations, we can split the excitations as:

\begin{equation}\label{excitation_splitting}
G = G^{ext} + G^{mat} \ ,
\end{equation}

or:

\begin{equation}\label{excitation_splitting_3+1}
\begin{split}
\Da &= \epsilon_0 \star E = \Da^{mat} + \Da^{ext} =  \Da^{ext} - P\left[ E, B \right] \\
\Ha &= \frac{1}{\mu_0}\star B = \Ha^{mat} + \Ha^{ext} =  \Ha^{ext} + M\left[ E, B \right] \ ,
\end{split}
\end{equation}

where the first equalities in both equations assume linear spacetime constitutive relations.

Differentiating $G^{ext}$ and using the Maxwell equations $\de G = J$ and \eqref{bounded_sources_potential}, we obtain the inhomogeneous Maxwell equations:

\begin{equation}\label{inhomogeneous_Maxwell_equations}
\de G^{ext} = J^{ext} \ ,
\end{equation}

or:

\begin{equation}\label{inhomogeneous_Maxwell_equations_3+1}
\begin{split}
\de \Da^{ext} &= \rho^{ext} \\
\de \Ha^{ext} - \partial_t \Da^{ext} &= j^{ext} \ .
\end{split}
\end{equation}

The external excitation 2-form $H^{ext}$ can be considered as an auxiliary field.

For linear media, we consider the following constitutive laws:

\begin{equation}\label{linear_constitutive_laws_magnetization_polarization}
\begin{split}
P &= \epsilon_0 \chi_E \star E \\
M &= \frac{1}{\mu_0} \chi_B \star B \ ,
\end{split}
\end{equation}

where $\chi_E$ is the electric susceptibility and $\chi_B$ the magnetic susceptibility. This leads to:

\begin{equation}\label{linear_constitutive_laws_excitations}
\begin{split}
D^{ext} &= \epsilon \star E \\
H^{ext} &= \frac{1}{\mu} \star B \ ,
\end{split}
\end{equation}

with the material constants defined as:

\begin{equation}\label{material_constants}
\begin{split}
\epsilon &\equiv \epsilon_0 \left(1 + \chi_E \right) \\
\mu &\equiv  \frac{\mu_0}{\left(1 + \chi_B \right)} \ .
\end{split}
\end{equation}

For linear dielectric media, we can express the constitutive relation as \cite{thompson2018covariant}:

\begin{equation}\label{Material_excitation_2_form}
\mathbf{G} = \star \mathbf{\chi} \mathbf{F} \ ,
\end{equation}

where $\mathbf{\chi}$ is a map from 2-forms to 2-forms, or in component form:

\begin{equation}
G_{\mu\nu} = \star_{\mu\nu}^{\alpha\beta} \chi_{\alpha\beta}^{\phantom{\alpha\beta}\sigma\rho} F_{\sigma\rho} \ .
\end{equation}

Finally, the constitutive relation \eqref{Material_excitation_2_form} yields the macroscopic Maxwell equations:

\begin{equation}\label{Macroscopic_Maxwell_equations}
\begin{split}
\de \mathbf{F} &= 0 \\
\de \mathbf{G} &= \mathbf{J}_{free} \ .
\end{split}
\end{equation}

This formulation provides a description of electromagnetic phenomena in material media, incorporating both microscopic and macroscopic aspects.

\subsection{Theory of the Hertzian Potential}

The Hertzian potential method offers an elegant and efficient approach to solving Maxwell's field equations in waveguides, particularly in flat spacetime \cite{collin1960field}. This powerful technique has been developed and refined over the years, with significant contributions from various researchers. \cite{nisbet1955hertzian} and \cite{mccrea1957hertzian} laid the groundwork for flat spacetime applications, with the latter introducing a covariant notation. \cite{cohen1974electromagnetic} provides a great treatment of the theory. \cite{mustafa1987hertz} further advanced the field by classifying Hertzian schemes using bivector potentials as eigenvectors of the Hodge duality operator, simplifying the solution of coupled equations. For those seeking practical applications, \cite{bouas2011hertz} offer valuable computational insights. Together, these works form a foundation for our exploration of Hertzian potentials in electromagnetic theory.

\subsubsection{Vector Formulation in Flat Spacetime}

In this section, we present a vector formulation of electromagnetic theory in flat spacetime using Hertzian potentials. We begin with the Maxwell equations in vacuum, using natural units (c = 1):

\begin{equation}\label{Maxwell_equations_flat_natural_units}
\begin{split}
\nabla \times \mathbf{E} + \partial_t \mathbf{B} = 0 \ , \\
\nabla \cdot \mathbf{B} = 0 \ , \\
\nabla \times \mathbf{B} - \partial_t \mathbf{E} = 0 \ , \\
\nabla \cdot \mathbf{E} = 0 \ ,
\end{split}
\end{equation}

Our goal is to express solutions to these equations in terms of the second derivatives of two bivector fields, $\mathbf{\Pi}_e$ and $\mathbf{\Pi}_m$, known as the \textbf{Hertzian potentials}.

The four-vector potential can be derived from first derivatives of the Hertzian bivector as follows \cite{cohen1974electromagnetic}:

\begin{equation}\label{Four_vector_0_from_Hertzian_potential}
\varphi = - \nabla \cdot \mathbf{\Pi}_e \ ,
\end{equation}

\begin{equation}\label{Four_vector_i_from_Hertzian_potential}
\mathbf{A} = \partial_t \mathbf{\Pi}_e + \nabla \times \mathbf{\Pi}_m \ .
\end{equation}

The electric field and magnetic induction can then be expressed in terms of these Hertzian potentials:

\begin{equation}
\begin{aligned}
\mathbf{E} &= \nabla \left(\nabla \cdot \mathbf{\Pi}_e \right) - \partial^2_t \mathbf{\Pi}_e - \nabla \times \partial_t \mathbf{\Pi}_m = - \nabla \times \partial_t \mathbf{\Pi}_m + \nabla \times \left( \nabla \times \mathbf{\Pi}_e\right) \ , \\
\mathbf{B} &=  \nabla \times \partial_t \mathbf{\Pi}_e + \nabla \times \left( \nabla \times \mathbf{\Pi}_m\right) = \nabla \left(\nabla \cdot \mathbf{\Pi}_m \right) - \partial^2_t \mathbf{\Pi}_m + \nabla \times \partial_t \mathbf{\Pi}_e \ .
\end{aligned}
\label{Hertzian_potential_fields}
\end{equation}

For these expressions to satisfy Maxwell's equations, the Hertzian bivectors must obey the following wave equations:

\begin{equation}\label{wave_equation_Hertzian_potential_vector_form}
\begin{split}
\square \mathbf{\Pi}_e &= 0 \ , \\
\square \mathbf{\Pi}_m &= 0 \ ,
\end{split}
\end{equation}

where $\square = -\partial_t^2 + \nabla^2$ is the D'Alembertian operator.

To introduce additional degrees of freedom, we can employ gauge transformations of the third kind \cite{nisbet1955hertzian}. These involve two four-vector gauge fields, $G = (g, \mathbf{G})$ and $W = (w, \mathbf{W})$:

\begin{equation}\label{gauge_four_vector_1}
\begin{split}
\mathbf{Q}_e &= \nabla \times \mathbf{G} \ , \\
\mathbf{Q}_m &= - \partial_t\mathbf{G} - \nabla g \ ,
\end{split}
\end{equation}

and:

\begin{equation}\label{gauge_four_vector_2}
\begin{split}
\mathbf{R}_e &= - \partial_t\mathbf{W} - \nabla w \ , \\
\mathbf{R}_m &= - \nabla \times \mathbf{W} \ .
\end{split}
\end{equation}

With these gauge transformations, the wave equations for the Hertzian potentials become:

\begin{equation}\label{gauged_wave_equation_Hertzian_potential_vector_form}
\begin{split}
\square \mathbf{\Pi}_e &= \mathbf{Q}_e + \mathbf{R}_e  \ , \\
\square \mathbf{\Pi}_m &= \mathbf{Q}_m +\mathbf{R}_m \ ,
\end{split}
\end{equation}

Consequently, the expressions for the electric field and magnetic induction are modified to:

\begin{equation}\label{electric_field_vector_Hertzian_potential}
\mathbf{E} = \mathbf{R}_e  + \nabla \left(\nabla \cdot \mathbf{\Pi}_e \right) - \partial^2_t \mathbf{\Pi}_e - \nabla \times \partial_t \mathbf{\Pi}_m = - \mathbf{Q}_e - \nabla \times \partial_t \mathbf{\Pi}_m + \nabla \times \left( \nabla \times \mathbf{\Pi}_e\right) \ ,
\end{equation}

and:

\begin{equation}\label{magnetic_induction_vector_Hertzian_potential}
\mathbf{B} = - \mathbf{R}_m  + \nabla \times \partial_t \mathbf{\Pi}_e + \nabla \times \left( \nabla \times \mathbf{\Pi}_m\right) =  \mathbf{Q}_m + \nabla \left(\nabla \cdot \mathbf{\Pi}_m \right) - \partial^2_t \mathbf{\Pi}_m + \nabla \times \partial_t \mathbf{\Pi}_e \ .
\end{equation}

These gauge-transformed expressions for $\mathbf{E}$ and $\mathbf{B}$ continue to satisfy the source-free Maxwell equations, providing a more general representation of electromagnetic fields in terms of Hertzian potentials.

\subsubsection{Covariant Formulation in Curved Spacetime}

The covariant formulation of electromagnetism using Hertzian potentials provides a powerful method for solving source-free problems in curved spacetimes. This approach extends the Hertz formalism to all curved spacetimes, offering a remarkable economy in representing the Maxwell field. By expressing the four-potential as the co-derivative of a 2-form $\mathbf{\Pi}$, we can construct solutions that automatically satisfy Maxwell's equations. The formalism also introduces a new type of gauge freedom, termed by Nisbet as "gauge transformations of the third kind," which adds flexibility to the solution process. This method is particularly useful as it reduces the arbitrary source-free Maxwell field to two scalar functions obeying a single separable second-order wave equation, reflecting the two degrees of freedom of a zero-rest-mass field.

We begin by writing the four-potential as the co-derivative of a 2-form $\mathbf{\Pi}$:

\begin{equation}
\mathbf{A} = \delta \mathbf{\Pi} \ ,
\end{equation}

If we impose the condition:

\begin{equation}\label{Vave_equation_Hertzian_bivector}
\square \mathbf{\Pi} = 0 \ ,
\end{equation}

then the Faraday 2-form can be expressed as:

\begin{equation}\label{Faraday_in_terms_Hertzian_bivector_vanishing_gauge_terms}
\mathbf{F} = \de\delta \mathbf{\Pi} = - \delta \de \mathbf{\Pi} \ ,
\end{equation}

By construction, this formulation satisfies the Maxwell equations \eqref{Maxwell_equations_coordinate_independent}.

To introduce additional flexibility, we can incorporate gauge degrees of freedom. Let $\mathbf{G}$ and $\mathbf{W}$ be arbitrary 1-forms. We define the 2-forms:

\begin{equation}\label{gauge_two_forms}
\begin{split}
\mathbf{Q} = \de \mathbf{G} \ , \\
\mathbf{R} = \star\de \mathbf{W} \ ,
\end{split}
\end{equation}

These can be used as gauge terms by introducing them as sources in the wave equation for the Hertzian potential:

\begin{equation}\label{gauged_wave_equation_for_Hertzian_potential}
\square \mathbf{\Pi} = \mathbf{Q} + \mathbf{R} = \de \mathbf{G} + \star\de \mathbf{W} \ ,
\end{equation}

Consequently, the gauge-transformed Faraday 2-form becomes:

\begin{equation}\label{gauge_transformed_Faraday}
\mathbf{F} = \de \delta \mathbf{\Pi} - \de \mathbf{G} = \star \de \mathbf{W} - \delta \de \mathbf{\Pi} \ .
\end{equation}

It can be shown that this gauge-transformed Faraday 2-form still satisfies the source-free Maxwell equations \eqref{Maxwell_equations_coordinate_independent}. These equations are valid for arbitrary curved spacetimes.

In the special case where $\mathbf{G} = \mathbf{W} = 0$, we can use \eqref{Faraday_in_terms_Hertzian_bivector_vanishing_gauge_terms} to express the action as:

\begin{equation}\label{EM_field_action_in terms_Hertzian_bivector_with_forms}
\begin{split}
S & = - \frac{1}{2}\int_\Ma \left(\de \delta \mathbf{\Pi} \wedge \star \de \delta \mathbf{\Pi} \right) = \frac{1}{2}\int_\Ma \left( \de \delta \mathbf{\Pi} \wedge \star \delta \de  \mathbf{\Pi} \right) =  - \frac{1}{2}\int_\Ma \left( \de \delta \mathbf{\Pi} \wedge \ \de \star \de  \mathbf{\Pi} \right)  \\
&=  \frac{1}{2}\int_\Ma \left( \de \delta \mathbf{\Pi} \wedge \ \de \delta \star  \mathbf{\Pi} \right) \ .
\end{split}
\end{equation}

\subsubsection{Separation of the Electromagnetic Degrees of Freedom}

To further simplify the problem, we can separate the electromagnetic degrees of freedom. We begin with the Maxwell equations for the Hertzian 2-form $\mathbf{\Pi}$:

\begin{equation}
(\de\delta + \delta \de)\mathbf{\Pi} = \de \mathbf{G} + \star \de \mathbf{W} \ ,
\end{equation}

We split the Hertzian potential into two parts:

\begin{equation}
\mathbf{\Pi} = \mathbf{\Pi}^{+} + \mathbf{\Pi}^{-} \ ,
\end{equation}

where $\mathbf{\Pi}^{+}$ and $\mathbf{\Pi}^{-}$ are contained in the subspaces of self-dual and anti-self-dual 2-forms, respectively:

\begin{equation}
\star \mathbf{\Pi}^{\pm} = \pm i \mathbf{\Pi}^{\pm} \ .
\end{equation}

Given arbitrary bases of these subspaces $\left\lbrace\theta^{\pm}_j \right\rbrace \ , j=1,..3$, we choose the gauge 1-forms to be:

\begin{equation}
\begin{split}
\mathbf{G} = \mathbf{G}^{+} + \mathbf{G}^{-} \\
\mathbf{M} = \mathbf{M}^{+} + \mathbf{M}^{-} \ ,
\end{split}
\end{equation}

with

\begin{equation}
\begin{split}
\mathbf{G}^{\pm} &= \Pi_j^{\pm}\delta\theta^{\pm}_j \mp i\tilde{G}^{\pm} \\
\mathbf{W}^{\pm} &= \Pi_j^{\pm}\star \de \theta^{\pm}_j + \tilde{W}^{\pm} \ .
\end{split}
\end{equation}

If we further assume $G^{\pm} = W^{\pm}$, this gauge leads to the equations:

\begin{equation}\label{separated_eq_1}
(i \pm \star) \de \left[\star \left(\de \Pi_j^{\pm}\wedge \theta^{\pm}_j \right)  - G^{\pm}\right] = 0 \ .
\end{equation}

There exists a basis on which:

\begin{equation}\label{basis}
\de \Pi_j^{\pm}\wedge \theta^{\pm}_j = 0 \quad \text{if} \quad j=1,2 \ ;
\end{equation}

Renaming $\Pi_3^{\pm} = \Pi^{\pm}$, $\theta^{\pm}_3 = \theta^{\pm}$, the equations \eqref{separated_eq_1} reduce to:

\begin{equation}\label{separated_eq_2}
(i \pm \star) \de \left[\star \left(\de \Pi^{\pm}\wedge \theta^{\pm} \right)  - G^{\pm}\right] = 0 \ .
\end{equation}

Using the relations:

\begin{equation}
\star (v\wedge \omega) = i_{\omega^{\flat}} \star v \ ,
\end{equation}

and the Cartan magic formula:

\begin{equation}
\de \left(i_X \omega\right) = \mathcal{L}_{X} \omega -i_X \de \omega \ ,
\end{equation}

the equations \eqref{separated_eq_2} become:

\begin{equation}\label{separated_eq_3}
(i \pm \star) \left[i_{\left(\theta^{\pm}\right)^{\flat}} \de\star\de \Pi^{\pm} - \left(\mathcal{L}_{\left(\theta^{\pm}\right)^{\flat}}\star\de \Pi^{\pm} + \de G^{\pm} \right)\right] = 0 \ .
\end{equation}

By fixing the gauge:

\begin{equation}
\de G^{\pm} = -\mathcal{L}_{\left(\theta^{\pm}\right)^{\flat}}\star\de \Pi^{\pm} \ ,
\end{equation}

the Maxwell equations reduces to the wave equation:

\begin{equation}
\de\star\de \Pi^{\pm} = 0 \rightarrow \delta \de \Pi^{\pm} = 0 \ .
\end{equation}

This formulation separates the Maxwell equations into two scalar equations, significantly simplifying the problem. The main challenge in practical applications is finding the appropriate basis \eqref{basis}.

\pagebreak
\section{Guided Waves in Static Spacetimes}\label{guided_waves_static_st_section}

This section explores electromagnetic wave propagation in waveguides within static curved spacetimes. We begin with a review of guided waves in flat spacetime using Hertzian potentials, then develop a model for waveguides in curved spacetimes. We derive the Maxwell equations and corresponding wave equations in a 3+1 formalism, and apply the Hertzian potential method to solve for guided waves. We analyze TE, TM, and TEM modes, examine their properties, and consider specific applications to the Schwarzschild spacetime.

\subsection{Guided Waves in Flat Spacetime with Hertzian Potentials}

We first summarize the results from \cite[Ch.5, p.329]{collin1960field} and then recover them using the covariant formalism. In our analysis, we take the $z$-axis as the axis of the guide and hence as the propagation direction. We assume harmonic time dependence $e^{i\omega t}$ for all field quantities.

\subsubsection{Vector Approach}

Consider the case where $\mathbf{R}_e = \mathbf{R}_m = \mathbf{Q}_e = \mathbf{Q}_m = 0$ and the Hertzian potentials are aligned with the $z$-axis: $\mathbf{\Pi}_e = \Pi_e \mathbf{e}_z$, $\mathbf{\Pi}_m = \Pi_m \mathbf{e}_z$. Under these conditions, we can express equations \eqref{electric_field_vector_Hertzian_potential} and \eqref{magnetic_induction_vector_Hertzian_potential} in Cartesian coordinates as:

\begin{equation}\label{electric_field_components_in_terms_Hertzian_potential}
\mathbf{E} = \left(\partial_x\partial_z \Pi_e - \partial_t\partial_y \Pi_m \right) \mathbf{e}_x + \left(\partial_y\partial_z \Pi_e + \partial_t \partial_x \Pi_m \right) \mathbf{e}_y + \left( \partial_z^2 \Pi_e - \partial_t^2 \Pi_e\right) \mathbf{e}_z \ ,
\end{equation}

and

\begin{equation}\label{magnetic_induction_components_in_terms_Hertzian_potential}
\begin{split}
\mathbf{B} &= \left(\partial_x\partial_z \Pi_m + \partial_t\partial_y \Pi_e \right) \mathbf{e}_x + \left(\partial_y\partial_z \Pi_m - \partial_t \partial_x \Pi_e \right) \mathbf{e}_y + \left(\partial_z^2 \Pi_m - \partial_t^2 \Pi_m \right) \mathbf{e}_z \\
&= \left(\partial_x\partial_z \Pi_m + \partial_t\partial_y \Pi_e \right) \mathbf{e}_x + \left(\partial_y\partial_z \Pi_m - \partial_t \partial_x \Pi_e \right) \mathbf{e}_y - \left(\partial_x^2 \Pi_m + \partial_y^2 \Pi_m \right) \mathbf{e}_z \ ,
\end{split}
\end{equation}

In the last equality of equation \eqref{magnetic_induction_components_in_terms_Hertzian_potential}, we have used the wave equation \eqref{wave_equation_Hertzian_potential_vector_form} to replace $\left(\partial_z^2  - \partial_t^2  \right)\Pi_m$ with $- \left(\partial_x^2  + \partial_y^2  \right)\Pi_m$.

Referring to the structure of the Faraday 2-form in Minkowski spacetime \eqref{Faraday_2_form_Minkowski}, we can express the Faraday 2-form corresponding to the fields \eqref{electric_field_components_in_terms_Hertzian_potential} and \eqref{magnetic_induction_components_in_terms_Hertzian_potential} as:

\begin{equation}\label{Faraday_2_form_in_terms_Hertzian_potential_flat}
\begin{split}
\mathbf{F} = &\left(\partial_t\partial_y\Pi_m  - \partial_x\partial_z\Pi_e\right) \de t \wedge \de x - \left(\partial_t\partial_x\Pi_m + \partial_y\partial_z\Pi_e\right) \de t \wedge \de y + \left(\partial_t^2\Pi_e - \partial_z^2\Pi_e \right) \de t \wedge \de z \\
&-\left(\partial_t\partial_y\Pi_e + \partial_x\partial_z \Pi_m \right) \star\de t \wedge \de x + \left(\partial_t\partial_x\Pi_e - \partial_y\partial_z\Pi_m \right) \star\de t \wedge \de y + \left(\partial_x^2\Pi_m + \partial_y^2\Pi_m \right) \star\de t \wedge \de z \ .
\end{split}
\end{equation}

\subsubsection{Rectangular Waveguides}

Rectangular waveguides support two types of modes: Transverse Electric (TE) and Transverse Magnetic (TM). We'll examine each type using the Hertzian potential approach.

\paragraph{TE Modes\newline}

TE modes are characterized by the absence of a longitudinal component of the electric field. We can derive the field components for these modes using a magnetic-type Hertzian potential with a single component along the guide's axis: $\mathbf{\Pi}_h = \Pi_h \mathbf{e}_z$. This potential satisfies the wave equation (see also \cite{nisbet1955hertzian}):

\begin{equation}\label{Wave_eq_magnetic_Hertzian_potential}
\square \mathbf{\Pi}_h = \nabla_s^2 \mathbf{\Pi}_h + k_0^2\mathbf{\Pi}_h = \nabla_s \left( \nabla_s \cdot \mathbf{\Pi}_h\right) - \nabla_s \times \nabla_s \times \mathbf{\Pi}_h + k_0^2\mathbf{\Pi}_h = 0 \ ,
\end{equation}

where $\nabla_s^2$ is the spatial Laplace-Beltrami operator (or vector Laplacian \cite{hirota1982vector}) and $k_0^2 = \omega^2 \mu_0 \epsilon_0 = \frac{4\pi^2}{\lambda_0^2}$, with $\lambda_0$ being the free-space wavelength of electromagnetic waves.

We can obtain the electric field and magnetic induction from equations \eqref{electric_field_vector_Hertzian_potential} and \eqref{magnetic_induction_vector_Hertzian_potential} respectively (see also \cite{nisbet1955hertzian}):

\begin{equation}\label{electric_field_TE_modes_flat}
\mathbf{E} = -i\omega \mu_0 \nabla_s \times \mathbf{\Pi}_h \ ,
\end{equation}

\begin{equation}\label{magnetic_induction_TE_modes_flat}
\mathbf{H} = k_0^2 \mathbf{\Pi}_h + \nabla_s \left( \nabla \cdot \mathbf{\Pi}_h\right) = \nabla_s \times \nabla_s \times \mathbf{\Pi}_h \ .
\end{equation}

The dominant TE mode in rectangular waveguides is the $TE_{10}$ (or $H_{10}$) mode \cite[Ch.5, p.329]{collin1960field}.

\paragraph{TM modes\newline}

For TM modes, we use an analogous approach but with an electric Hertzian potential $\mathbf{\Pi}_e = \Pi_e \mathbf{e}_z$. This potential also satisfies a wave equation:

\begin{equation}\label{Wave_eq_electric_Hertzian_potential}
\square \mathbf{\Pi}_e = \nabla_s^2 \mathbf{\Pi}_e + k_0^2\mathbf{\Pi}_e = \nabla_s \left( \nabla_s \cdot \mathbf{\Pi}_e\right) - \nabla_s \times \nabla_s \times \mathbf{\Pi}_e + k_0^2\mathbf{\Pi}_e = 0 \ ,
\end{equation}

The electric field and magnetic induction for TM modes can be derived as follows (see also \cite{nisbet1955hertzian}):

\begin{equation}\label{electric_field_TM_modes_flat}
\mathbf{E} = k_0^2 \mathbf{\Pi}_e + \nabla_s\left( \nabla_s \cdot \mathbf{\Pi}_e\right) =  \nabla_s \times \nabla_s \times \mathbf{\Pi}_e \ ,
\end{equation}

\begin{equation}\label{magnetic_induction_TM_modes_flat}
\mathbf{H} = i\omega \epsilon_0 \nabla_s \times \mathbf{\Pi}_e \ .
\end{equation}

The dominant TM mode in rectangular waveguides is the $TM_{11}$ (or $E_{11}$) mode \cite[Ch.5, p.329]{collin1960field}.

\subsubsection{Covariant Approach}

We now demonstrate how to obtain the same results using covariant notation, following the approach in \cite{bouas2011hertz}.

Consider the Minkowski metric:

\begin{equation}\label{Minkowski_metric}
\de s^2 = \eta_{\mu \nu}\de x^\mu \de x^\nu = - c^2\de t^2 + \delta_{ij}\de x^i\de x^j \ ,
\end{equation}

where $i,j = x, y, z$.

We choose the Hertzian potential to be:

\begin{equation}\label{Hertzian_potential_ansatz_flat}
\begin{split}
\mathbf{\Pi} &= \Pi_e(t,x,y,z) \de t \wedge \de z - \Pi_m(t,x,y,z)\star\de t \wedge \de z \\
&= \Pi_e(t,x,y,z) \de t \wedge \de z + \Pi_m(t,x,y,z)\de x \wedge \de y \ .
\end{split}
\end{equation}

Applying the Laplace-de Rham operator:

\begin{equation}\label{de_Rham_flat}
\square \mathbf{\Pi} = \square \Pi_e(t,x,y,z) \de t \wedge \de z + \square\Pi_m(t,x,y,z)\de x \wedge \de y \ ,
\end{equation}

where, on functions, $\square$ is minus the usual D'Alembertian:

\begin{equation}\label{Laplacian_functions_flat}
\square f(t,x,y,z) = \left( \partial_t^2 - \nabla_s^2 \right) f(t,x,y,z) \ .
\end{equation}

In this case, we don't need non-vanishing gauge terms, so we set $\mathbf{G} = \mathbf{W} = 0$. The wave equations for the components of the Hertzian potential then simplify to:

\begin{equation}
\begin{split}
\square \Pi_m (t,x,y,z)= 0 \ , \\
\square \Pi_e (t,x,y,z)= 0 \ ,
\end{split}
\end{equation}

which coincides with equations \eqref{Wave_eq_electric_Hertzian_potential} and \eqref{Wave_eq_magnetic_Hertzian_potential}.

Once we have a solution to these wave equations, we use equation \eqref{gauge_transformed_Faraday} to find the Faraday 2-form:

\begin{equation}\label{Faraday_2_form_in_terms_Hertzian_potential_flat_Maple}
\begin{split}
\mathbf{F} &= \de \delta \mathbf{\Pi} - \de \mathbf{G} = \de \delta \mathbf{\Pi}  \\
& = \left(\partial_t \partial_y \Pi_m - \partial_x \partial_z \Pi_e \right) \de t \wedge \de x - \left(\partial_t \partial_x \Pi_m +  \partial_y \partial_z \Pi_e \right) \de t \wedge \de y + \left(\partial_t^2 \Pi_e - \partial_z^2 \Pi_e  \right) \de t \wedge \de z \\
& \left(\partial_t \partial_y \Pi_e + \partial_x \partial_z \Pi_m \right) \de y \wedge \de z  + \left(\partial_t \partial_x \Pi_e -  \partial_y \partial_z \Pi_m \right) \de x \wedge \de z  -  \left(\partial_x^2 \Pi_m + \partial_y^2 \Pi_m \right) \de x \wedge \de y \\
& = \left(\partial_t \partial_y \Pi_m - \partial_x \partial_z \Pi_e \right) \de t \wedge \de x - \left(\partial_t \partial_x \Pi_m +  \partial_y \partial_z \Pi_e \right) \de t \wedge \de y + \left(\partial_t^2 \Pi_e - \partial_z^2 \Pi_e  \right) \de t \wedge \de z \\
& - \left(\partial_t \partial_y \Pi_e + \partial_x \partial_z \Pi_m \right)\star \de t \wedge \de x  + \left(\partial_t \partial_x \Pi_e -  \partial_y \partial_z \Pi_m \right)\star \de t \wedge \de y  +  \left(\partial_x^2 \Pi_m + \partial_y^2 \Pi_m \right) \star \de t \wedge \de z  \ ,
\end{split}
\end{equation}

This result exactly matches equation \eqref{Faraday_2_form_in_terms_Hertzian_potential_flat}. 

We can thus identify the Hertzian 2-form components $\Pi_e$ and $\Pi_m$ in both formalisms. From equation \eqref{Hertzian_potential_ansatz_flat}, we can obtain:

\begin{itemize}
    \item TE modes (equations \eqref{electric_field_TE_modes_flat} and \eqref{magnetic_induction_TE_modes_flat}) by setting $\Pi_m \neq 0, \Pi_e = 0$.
    \item TM modes (equations \eqref{electric_field_TM_modes_flat} and \eqref{magnetic_induction_TM_modes_flat}) by setting $\Pi_m = 0, \Pi_e \neq 0$.
\end{itemize}

This covariant approach provides a more geometrical understanding of the electromagnetic fields in waveguides and demonstrates the equivalence between the vector and covariant formulations.

\subsection{Waveguide Model and Induced Metric}

\subsubsection{Waveguide Model}

Our waveguide model builds upon the abstract differential forms approach developed for flat spacetime \cite{tucker2007effects} and extended to curved spacetime \cite{burton2005twisted}. Following Burton et al. \cite{burton2005twisted}, we consider 'wavetubes' - cavities whose transversal section is small compared to both their length and the scale of gravitational field variation. This allows us to assume a locally flat spacetime throughout the cross-section, with curved spacetime effects primarily affecting the axial dimension.
We describe the wavetube as a time-dependent curve $\Gamma(t)$ within the spacetime, defined on constant time hypersurfaces. This curve represents the waveguide's axis, chosen based on symmetry considerations. For simplicity, we focus on time-independent waveguides in this work.

The boundary of the wavetube is defined as a hypersurface (perfectly conducting, in the case of a 'wavetube' as those considered in \cite{burton2005twisted}):

\begin{equation}\label{wavetube_boundary}
\Ba \equiv \left\lbrace p \in \Ma \mid f(p) = 0, \ \de f \ \text{spacelike} \right\rbrace \ .
\end{equation}

The topology of the wavetube is hence $\mathbb{R}^2 \times \Da$, where $\Da$ is the 2-disc submanifold associated with the transversal section of the waveguide. This allows us to write the boundary condition for the Faraday 2-form as:

\begin{equation}\label{boundary_condition_Faraday_2_form}
\de f \wedge \mathbf{F}  = 0 \quad \forall p \in \Ba \ .
\end{equation}

This formulation provides the framework for analyzing electromagnetic wave propagation in curved spacetime waveguides.

\subsubsection{Metric Derivation: Schwarzschild Example}

To illustrate our approach, let's consider the Schwarzschild spacetime\footnote{The Schwarzschild spacetime is given by \cite{Will_Gravity}:

\begin{equation}\label{Schwarzschild_metric}
\de s^2 = - \left(1 - \frac{R_s}{r} \right) \de (c t)^2 + \left(1 - \frac{R_s}{r} \right)^{-1} \de r^2 + r^2 \de \Omega^2 \ ,
\end{equation}

with

\begin{equation}\label{Schwarzschild_radius}
R_s \equiv \frac{2GM}{c^2} \ .
\end{equation}
}. 

We can write the metric as:

\begin{equation}
\label{Schwarzschild_metric_2}
\de s^2 = - \left(1 - \frac{R_s}{r} \right) \de (ct)^2  + \gamma_{ij}\de x^i \de x^j \ ,
\end{equation}

where we have introduced the spatial metric $\gamma_{ij}\de x^i \de x^j$, which in spherical coordinates reads:

\begin{equation}\label{spatial_metric}
\gamma_{ij}\de x^i \de x^j =  \left(1 - \frac{R_s}{r} \right)^{-1} \de r^2 + r^2 \de \Omega^2 \ . 
\end{equation}

To parameterize the waveguide $\Gamma$ in terms of its proper length $l$, we consider an arbitrary parameterization $\mathbf{x}(\lambda)$ of the curve. The line element on $\Gamma$ is then:

\begin{equation}\label{differential_lenght}
\de l =  \sqrt{\gamma_{ij}(\lambda)\frac{\de x^i(\lambda)}{\de \lambda}\frac{\de x^j(\lambda)}{\de \lambda}}\de \lambda \ ,
\end{equation}

and its proper lenght is given by:

\begin{equation}\label{proper_lenght}
l_{\Gamma} = \int_\Gamma \de l \ .
\end{equation}

In terms of this lenght, the line element along the waveguide can be written as:

\begin{equation}
\label{Schwarzschild_metric_3}
\de s^2 = - \left(1 - \frac{R_s}{r(l)} \right) \de (ct)^2  + \de l ^2 \ ,
\end{equation}

or in abstract form:

\begin{equation}\label{Schwarzschild_metric_abstract}
\de s^2 = - f(l) \de (ct)^2 + \de l ^2 \ .
\end{equation}

Neglecting the spatial curvature effects on the section of the waveguide, as the transversal section is narrow, we can introduce a pair of space-like coordinates to account for these dimensions, and consider a metric of the form:

\begin{equation}
\label{general_metric_abstract}
\de s^2 = - f(l) \de (ct)^2 + \de l ^2 + g^{\perp}_{mn}\de x^m \de x^n \ ,
\end{equation}

where $g^{\perp}_{mn}$ is the metric for $\mathbb{R}^2$ in some coordinates.

\paragraph{Radial Propagation\newline}

For pure radial propagation, we have $\de \Omega = 0$, and the proper spatial length is given by:

\begin{equation}\label{proper_exact_radial_Schwarzschild}
l = \int^r_{r_0} \de r' \frac{1}{\sqrt{1 - \frac{R_s}{r'}}} \ .
\end{equation}

This indefinite integral has analytic solution, but in the case in which the waveguide is far from the Schwarzschild radius\footnote{The Schwarzschild radius of the Earth is $R_{s,earth} \simeq 9 \cdot 10^{-3} \ \text{m}$,  while the radius of the Earth is $R_{earth} \simeq 6.371 \cdot 10^3 \ \text{m}$. For the Sun, the Schwarzschild radius is approximately $R_{s,sun} \simeq 3 \cdot 10^{6} \ \text{m}$, and its radius $R_{sun} \simeq 6.957 \cdot 10^8 \ \text{m}$. In both cases, for waveguides placed outside the planetary/stellar object, the approximation \eqref{weak_field_condition} holds.}, i.e when:

\begin{equation}\label{weak_field_condition}
r >> R_s \qquad \forall r \in \Gamma \ ,
\end{equation}

we can perform the approximation \footnote{If |x|<1:

\begin{equation}\label{Taylor_series_1}
\frac{1}{\sqrt{1 + x}} = 1 - \frac{1}{2}x + \frac{3}{8}x^2 - \frac{5}{16}x^3 + \cdots \ .
\end{equation}
}:

\begin{equation}\label{proper_approx_radial_Schwarzschild}
l \simeq \int^r_{r_0} \de r' \left( 1 + \frac{1}{2}\frac{r_s}{r'}\right) = r + \frac{r_s}{2}ln\left(\frac{r}{r_0}\right) - r_0 \ .
\end{equation}

\subsubsection{Metric Properties}\label{secc_metric}

Motivated by the result \eqref{general_metric_abstract}, we study the propagation of the electromagnetic field in the geometry given by the line element:

\begin{equation}\label{Metric}
    ds^2 = - f(s)dt^2 + ds^2 + du^2 + dv^2 \ .
\end{equation}

Here, $s$ represents the axial coordinate (spatial proper length along the waveguide axis), while $u$ and $v$ are the transversal coordinates in a flat space (assuming a small transversal section). The component $g_{tt} (s) = - f(s)$ of the metric represents the induced geometry along the waveguide.

Due to the symmetries of the metric \eqref{Metric}, generated by the Killing vectors $\partial_t$, $\partial_u$, and $\partial_v$, we expand the fields in 'eigenstates' of these generators, i.e., in harmonic waves:

\begin{equation}\label{basic_ansatz}
\phi(x) \propto e^{-i\omega t}e^{ik_u u}e^{ik_v v} \ .
\end{equation}

For this geometry, the non-vanishing Christoffel symbols are given by (modulo its symmetric counterparts in the lower indices):

\begin{equation}\label{Christoffel_symbols}
    \Gamma^s_{\phantom{s}tt} = \frac{1}{2} f'(s)  \ , \qquad
    \Gamma^t_{\phantom{t}st} = \frac{1}{2} \frac{d}{ds} \lbrace ln[f(s)] \rbrace \ ,
\end{equation}

where $()' \equiv \frac{d}{ds}()$.
 
The non-vanishing components of the Ricci tensor are:
 
 \begin{equation}
      R_{tt} = \frac{1}{2}\left[f''(s) - \frac{1}{2}\frac{(f'(s))^2}{f(s)}\right] \ , \qquad
      R_{ss} = \frac{1}{4(f(s))^2}\left[(f'(s))^2 - 2f(s)f''(s) \right] \ .
 \end{equation}

These components yield a vacuum constraint on $f(s)$ (imposing the metric to be Ricci flat, i.e., $R_{\mu\nu} = 0$):
 
 \begin{equation}\label{Ricci_flat_condition}
     f''(s) = \frac{1}{2}\frac{(f'(s))^2}{f(s)} \ .
 \end{equation}
 
\paragraph{Laplace-de Rham Operator on Functions\newline}
 
The Laplace-de Rham operator (minus Laplacian) on functions is:

\begin{equation}\label{Laplacian_functions_spatially_flat}
\square f(t,x,y,z) = \left( \frac{1}{f(z)}\partial_t^2 - \nabla_s^2  - \frac{1}{2}\frac{f'(z)}{f(z)}\partial_z\right)f(t,x,y,z)  \ .
\end{equation}
 
\paragraph{Hodge Star Action on Basis p-forms\newline}
 
On 0-forms:
 
\begin{equation}\label{Hodge_star_on_0_forms}
 \star f(x) = f(x) \sqrt{f(z)} \de t \wedge \de x \wedge \de y \wedge \de z \ .
\end{equation}

On 1-forms: 
 
\begin{equation}\label{Hodge_star_on_1_forms}
\begin{split}
 \star \de t &= -\frac{1}{\sqrt{f(z)}} \de x \wedge \de y \wedge \de z  \\
\star  \de x &= -\sqrt{f(z)} \de t \wedge \de y \wedge \de z \\
\star  \de y &= \sqrt{f(z)} \de t \wedge \de x \wedge \de z \\
\star  \de z &= -\sqrt{f(z)} \de t \wedge \de x \wedge \de y \ . 
\end{split}
\end{equation} 

On 2-forms:

\begin{equation}\label{Hodge_star_on_2_forms}
\begin{split}
 \star \de t \wedge \de x &= -\frac{1}{\sqrt{f(z)}} \de y \wedge \de z  \\
\star  \de t \wedge \de y &= \frac{1}{\sqrt{f(z)}} \de x \wedge \de z  \\
\star  \de t \wedge \de z &= -\frac{1}{\sqrt{f(z)}} \de x \wedge \de y \\
\star  \de x \wedge \de y &= \sqrt{f(z)} \de t \wedge \de z  \\ 
\star  \de x \wedge \de z &= -\sqrt{f(z)} \de t \wedge \de y \\ 
\star  \de y \wedge \de z &= \sqrt{f(z)} \de t \wedge \de x  \ . 
\end{split}
\end{equation} 

On 3-forms:

\begin{equation}\label{Hodge_star_on_3_forms}
\begin{split}
 \star \de t \wedge \de x \wedge \de y &= -\frac{1}{\sqrt{f(z)}} \de z   \\
\star  \de t \wedge \de x \wedge \de z &= \frac{1}{\sqrt{f(z)}} \de y \\
\star  \de t \wedge \de y \wedge \de z &= -\frac{1}{\sqrt{f(z)}} \de x  \\
\star  \de x \wedge \de y \wedge \de z &= -\sqrt{f(z)} \de t    \ . 
\end{split}
\end{equation} 

On 4-forms:

\begin{equation}\label{Hodge_star_on_4_forms}
 \star \de t \wedge \de x \wedge \de y  \wedge \de z = -\frac{1}{\sqrt{f(z)}}   \ . 
\end{equation} 
  
\subsection{3+1 Field Theory. Maxwell and Wave Equations}

This section develops the 3+1 formalism for electromagnetic fields in static curved spacetimes. We begin by deriving the Maxwell equations in this context, paying particular attention to the effects of spacetime curvature. We then obtain wave equations for both the electric field and magnetic induction, considering their axial and transverse components separately. The analysis ends with a demonstration of the existence of TE and TM modes in arbitrary static spacetimes, generalizing the familiar concepts from flat spacetime electrodynamics.

\subsubsection{Maxwell Equations}

We begin by particularizing the inhomogeneous equations \eqref{Maxwell_equations_vacuum_inhomogeneous}. Using \eqref{Christoffel_symbols}, this reduces to:

\begin{equation}\label{Maxwell_equations_vacuum_inhomogeneous_particularization}
    \nabla_\mu F^{\mu \nu} = \partial_\mu F^{\mu \nu} + \Gamma^t_{\phantom{t}r t}F^{s \nu} = \partial_\mu F^{\mu \nu} + \frac{1}{2}\left(\frac{f'(s)}{f(s)} \right)F^{s \nu} \ .
\end{equation}

Lowering the indices of $F^{\mu \nu}$ in this equation, we obtain\footnote{The derivation of the Maxwell and wave equations in terms of the electric field $\mathbf{E}$ and the magnetic induction $\mathbf{B}$ parallels that in \cite{Cabral_electrodynamics}}:

\begin{equation}
    g^{\alpha \mu} g^{\beta \nu}\partial_\mu F_{\alpha\beta} + \left[\partial_s(g^{\nu \beta}) + \frac{1}{2}\left(\frac{f'(s)}{f(s)} \right)g^{\nu \beta}\right]F_{s\beta} = 0 \ .
\end{equation}

Using the definitions \eqref{Electric_field_definition} and \eqref{Magnetic_induction_definition}, and considering that we are working in a spacetime with a diagonal metric, we obtain:

\begin{equation}\label{Maxwell_equations_in_terms_electric_magnetic_fields}
\begin{split}
    & -\frac{1}{f(s)}g^{k\nu}\partial_tE_k - g^{ji}g^{t\nu}\partial_iE_j - \epsilon_{jki}g^{jl}g^{k\nu}\partial_lB^i \\ 
    & - \left[\partial_s(g^{\nu t}) + \frac{1}{2}\left( \frac{f'(s)}{f(s)}\right)g^{\nu t}\right]E_s - \left[\frac{1}{2}\left( \frac{f'(s)}{f(s)}\right)g^{\nu k}\right]\epsilon_{ski}B^i = 0 \ .
\end{split}
\end{equation}

From \eqref{Maxwell_equations_in_terms_electric_magnetic_fields}, we derive the Gauss law by setting $\nu = t$:

\begin{equation}\label{Gauss_law}
\nabla \cdot \mathbf{E} - \frac{1}{2}\left( \frac{f'(s)}{f(s)}\right) E_s = 0 \ .
\end{equation}

The Maxwell-Ampère law corresponds to $\nu = s,u,v$:

\begin{equation}\label{Maxwell_Ampere_law}
    -\partial_t \mathbf{E} + f(s) \left(\nabla \times \mathbf{B}\right) + \frac{1}{2}\left(\nabla f(s) \times \mathbf{B} \right) = 0 \ .
\end{equation}

\subsubsection{Wave Equations for Electric Field and Magnetic Induction}

We derive the wave equation for the electric field by differentiating the Maxwell-Ampère law \eqref{Maxwell_Ampere_law} with respect to time, using the Faraday law \eqref{Faraday_law} to eliminate terms involving the magnetic induction, and replacing the divergences of the electric field using the Gauss law \eqref{Gauss_law}. The result is:

\begin{equation}\label{Wave_equation_electric_field}
    \left[-\frac{1}{f(s)} \partial_t^2 + \nabla^2 \right] \mathbf{E} - \frac{1}{2}\nabla\left[\frac{f'(s)}{f(s)}\right]E_s + \frac{1}{2}\left( \frac{f'(s)}{f(s)}\right)\left[\partial_s \mathbf{E} - 2 \nabla E_s\right] = 0 \ ,
\end{equation}

where $\nabla^2 = \sum_i \partial_i^2$ is the spatial Laplacian operator.

The wave equation for the axial component $E_s$ is:

\begin{equation}\label{Wave_equation_electric_field_axial}
     \left[-\frac{1}{f(s)} \partial_t^2 + \nabla^2 \right] E_s - \frac{1}{2}\left(\frac{f'(s)}{f(s)}\right)\partial_s E_s - \frac{1}{2}\frac{d}{ds}\left(\frac{f'(s)}{f(s)}\right) E_s = 0 \ .
\end{equation}

Inserting the ansatz:

\begin{equation}\label{Axial_electric_field_ansatz}
E_s(x) = \phi_E(s)e^{i \mathbf{k}_\perp \cdot \mathbf{v}_\perp}e^{-i\omega t} \ ,
\end{equation}

where we have defined $\mathbf{k}_\perp = (k_u, k_v)^T$ and $\mathbf{v}_\perp = (u, v)^T$, we get the wave equation for the axial part $\phi_E(s)$:

\begin{equation}\label{Wave_equation_axial_part_electric field}
\phi_E''(s) - \frac{1}{2}\frac{d}{ds}\left[\left(\frac{f'(s)}{f(s)} \right)\phi_E(s)\right] + \left(\frac{\omega^2}{f(s)} - \mathbf{k}_\perp^2 \right) \phi_E(s) = 0 \ .
\end{equation}

Alternatively, making use of the Ricci flat condition for the metric \eqref{Ricci_flat_condition} the wave equation takes the form:

\begin{equation}\label{Wave_equation_axial_part_electric field_vacuum}
\phi_E''(s) - \frac{1}{2}\left(\frac{f'(s)}{f(s)} \right)\phi'_E(s) + \left[\frac{\omega^2}{f(s)} - \mathbf{k}_\perp^2 + \frac{1}{4}\left(\frac{f'(s)}{f(s)}\right)^2\right] \phi_E(s) = 0 \ .
\end{equation}

The wave equation for the magnetic induction appears if we take the temporal derivative of the Faraday law, then we use the Maxwell-Ampère law to get rid of the temporal derivative of the electric field and finally we cancel out the divergences of the magnetic induction due to the (magnetic) Gauss law. The result is:

\begin{equation}\label{Wave_equation_magnetic_induction}
    \left[-\frac{1}{f(s)} \partial_t^2 + \nabla^2 \right] \mathbf{B} = \left(\frac{f'(s)}{f(s)}\right)\left(\nabla B^s - \frac{3}{2}\partial_s \mathbf{B} \right) + \frac{1}{2}\left(\frac{f''(s)}{f(s)}\right)\left(\hat{\mathbf{a}} \cdot \mathbf{B} - \mathbf{B}\right) \ ,
\end{equation}

where $\hat{\mathbf{a}} \equiv (1, 0, 0)$ is an unit axial vector.

The axial component satisfies:

\begin{equation}\label{Wave_equation_magnetic_induction_axial}
    \left[-\frac{1}{f(s)} \partial_t^2 + \nabla^2 \right] B^s = -\frac{1}{2}\left(\frac{f'(s)}{f(s)} \right)\partial_s B^s \ .
\end{equation}

Writing the magnetic induction as:

\begin{equation}\label{Axial_magnetic_induction_ansatz}
B_s(x) = \phi_B(s)e^{i \mathbf{k}_\perp \cdot \mathbf{v}_\perp}e^{-i\omega t} \ ,
\end{equation}

the wave equation for the axial dependence of the axial magnetic induction becomes:

\begin{equation}\label{Wave_equation_axial_part_magnetic_induction}
\phi_B''(s) + \frac{1}{2}\left(\frac{f'(s)}{f(s)} \right)\phi_B'(s) + \left[\frac{\omega^2}{f(s)} - \mathbf{k}_\perp^2 \right] \phi_B(s) = 0 \ .
\end{equation}

\subsubsection{Guided Waves in Static Spacetimes: 3+1 Approach}

We now demonstrate the split into two fundamental sets of modes of the Maxwell equations inside the guide: the TE and TM modes. We generalize the approach followed in \cite[ch.21]{Garg_EM_nutshell} for the flat case.

Taking the cross product of a unit axial vector $\hat{\mathbf{a}}$ with the Faraday law:

\begin{equation}
    \hat{\mathbf{a}} \times \left( \partial_t \mathbf{B} + \nabla \times \mathbf{E} \right) = 0 \ ,
\end{equation}

we obtain:

\begin{equation}\label{Transversal_Faraday_law}
    -\partial_s \mathbf{E}_\perp + \partial_t (\hat{\mathbf{a}} \times \mathbf{B}_\perp) = - \nabla_\perp E_s \ ,
\end{equation}

where $\mathbf{E}_\perp$ and  $\mathbf{B}_\perp$ are the transversal parts of the fields and $\nabla_\perp$ the transversal gradient.

Taking the cross product of the Maxwell-Ampère law with $\hat{\mathbf{a}}$ twice:

\begin{equation}
     \hat{\mathbf{a}} \times \hat{\mathbf{a}} \times \left[ -\partial_t \mathbf{E} + f(s) \left(\nabla \times \mathbf{B}\right) + \frac{1}{2}\left(\nabla f(s) \times \mathbf{B} \right) \right] = 0 \ ,
\end{equation}

yields:

\begin{equation}\label{Transversal_Maxwell_Ampere_law}
    -\frac{1}{f(s)}\partial_t\mathbf{E}_\perp +  \partial_s \left(\hat{\mathbf{a}} \times \mathbf{B}_\perp\right) + \frac{1}{2}\left(\frac{f'(s)}{f(s)} \right)\left(\hat{\mathbf{a}} \times \mathbf{B}_\perp \right) =  \hat{\mathbf{a}} \times \nabla_\perp B_s \ .
\end{equation}

Introducing harmonic time dependence, equations \eqref{Transversal_Faraday_law} and \eqref{Transversal_Maxwell_Ampere_law} become:

\begin{equation}\label{Transversal_Faraday_law_harmonic}
    -\partial_s \mathbf{E}_\perp - i\omega (\hat{\mathbf{a}} \times \mathbf{B}_\perp) = - \nabla_\perp E_s \ ,
\end{equation}

\begin{equation}\label{Transversal_Maxwell_Ampere_law_harmonic}
    i\frac{\omega}{f(s)}\mathbf{E}_\perp +  \partial_s \left(\hat{\mathbf{a}} \times \mathbf{B}_\perp\right) + \frac{1}{2}\left(\frac{f'(s)}{f(s)} \right)\left(\hat{\mathbf{a}} \times \mathbf{B}_\perp \right) =  \hat{\mathbf{a}} \times \nabla_\perp B_s \ .
\end{equation}

Defining the operators:

\begin{equation}\label{Operators}
\hat{A} = - \partial_s \ , \qquad \hat{D} = -\hat{A} + \frac{1}{2}\left(\frac{f'(s)}{f(s)} \right) \ ,
\end{equation}

and the functions:

\begin{equation}\label{functions}
B = -i\omega \ , \qquad C = -\frac{B}{f(s)} \ ,
\end{equation}

the system of equations \eqref{Transversal_Faraday_law_harmonic} and \eqref{Transversal_Maxwell_Ampere_law_harmonic} can be written compactly as:

\begin{equation}
\begin{split}
\hat{A} \mathbf{E}_\perp + B(\hat{\mathbf{a}} \times \mathbf{B}_\perp) = - \nabla_\perp E_s \ , \\
C\mathbf{E}_\perp +  \hat{D}\left(\hat{\mathbf{a}} \times \mathbf{B}_\perp \right) =  \hat{\mathbf{a}} \times \nabla_\perp B_s \ .
\end{split}
\end{equation}

Finally, we decouple the system into one equation for $\hat{\mathbf{a}} \times \mathbf{B}_\perp$ and another for $\mathbf{E}_\perp$, both in terms of the axial components of the field:

\begin{equation}
\left[i \frac{f'(s)}{\omega}\hat{D} - \frac{\hat{A}\hat{D}}{C} - B\right]\hat{\mathbf{a}} \times \mathbf{B}_\perp = \nabla_\perp E_s + \left[i \frac{f'(s)}{\omega} + \frac{\hat{A}}{C}\right]\hat{\mathbf{a}} \times \nabla_\perp B_s  \ ,
\end{equation}

\begin{equation}
\left[C - \frac{\hat{D}\hat{A}}{B} \right]\mathbf{E}_\perp = \frac{\hat{D}}{B}\nabla_\perp E_s + \hat{\mathbf{a}} \times \nabla_\perp B_s  \ .
\end{equation}

These equations demonstrate the existence of TE and TM modes for arbitrary stationary spacetimes.

\subsection{Guided Waves in Static Spacetimes with Hertzian Potentials}

This section extends the Hertzian potential formalism to analyze guided electromagnetic waves in static curved spacetimes. We begin by considering a general static metric and develop the wave equations for Hertzian potentials in this context. Two approaches are presented: an initial attempt and a refined solution using an orthonormal coframe. We then explore the implications for material media and derive the equations for TE and TM modes. Special attention is given to the axial part of the solutions, cutoff frequencies, and guide wavelengths. We also examine TEM modes and their unique properties in curved spacetime. Finally, we apply this formalism to the specific case of the Schwarzschild spacetime, providing insights into radial propagation and TEM modes in strong gravitational fields.

Let us consider the propagation in the geometry \eqref{Metric}

\begin{equation}\label{Metric_x_y_z}
    \de s^2 = - f(z)\de t^2 + \de x^2 + \de y^2 + \de z^2 \ .
\end{equation}

\subsubsection{Initial Formulation in Coordinate Basis}

Following the approach developed for the flat spacetime problem, we consider the Hertzian potential in the 2-form basis:

\begin{equation}\label{Hertzian_potential_ansatz_spatially_flat}
\begin{split}
\mathbf{\Pi} &= \Pi_e(t,x,y,z) \de t \wedge \de z - \Pi_m(t,x,y,z)\star\de t \wedge \de z \\
&= \Pi_e(t,x,y,z) \de t \wedge \de z + \frac{1}{\sqrt{f(z)}}\Pi_m(t,x,y,z)\de x \wedge \de y \ . 
\end{split}
\end{equation}

Using the expression \eqref{Laplacian_functions_spatially_flat}, we can write the Laplace-de Rham operator acting on the Hertzian potential in terms of its action on the components of the Hertzian potential 2-form:

\begin{equation}\label{de_Rham__spatially_flat}
\begin{split}
\square \mathbf{\Pi} = &\left\lbrace \square \Pi_e + \frac{1}{2}\left( \frac{f'}{f}\right)\Pi_e' + \frac{1}{2}\partial_z\left[\left( \frac{f'}{f}\right)\Pi_e \right]\right\rbrace \de t \wedge \de z \\
& - \left\lbrace \square\Pi_m + \frac{1}{2}\left( \frac{f'}{f}\right)\Pi_m' + \frac{1}{2}\partial_z\left[\left( \frac{f'}{f}\right)\Pi_m \right] \right\rbrace \star\de t \wedge \de z \ .
\end{split}
\end{equation}

If we choose the gauge terms to be zero, $\mathbf{G} = \mathbf{W} = 0$, we obtain from $\square \mathbf{\Pi} = 0$ the same equation of motion for both components $\Pi_m$ and $\Pi_e$. We write this equation in terms of a field $\phi(t, x, y ,z)$ that represents both components:

\begin{equation}\label{Wave_equation_components_Hertzian_potential_1}
\square \phi + \frac{1}{2}\left( \frac{f'}{f}\right)\partial_z\phi + \frac{1}{2}\partial_z\left[\left( \frac{f'}{f}\right)\phi \right] = 0 \ ,
\end{equation}

which can also be written, using \eqref{Laplacian_functions_spatially_flat}, as:

\begin{equation}\label{Wave_equation_components_Hertzian_potential_2}
\left[ \frac{1}{f(z)}\partial_t^2 - \nabla_s^2 + \frac{1}{2}\partial_z\left( \frac{f'}{f}\right) + \frac{1}{2}\left( \frac{f'}{f}\right) \partial_z \right]\phi(t,x,y,z) = 0 \ ,
\end{equation}

This is the same equation as the wave equation for the axial component of the electric field \eqref{Wave_equation_electric_field_axial}.

Analogous to our approach for the electromagnetic field, we insert the ansatz:

\begin{equation}\label{Ansatz_Hertzian_potential_components}
\phi(x) = \phi_a(z)e^{i \mathbf{k}_\perp \cdot \mathbf{v}_\perp}e^{-i\omega t} \ ,
\end{equation}

to obtain the equation for the axial factor $\phi_a(z)$:

\begin{equation}\label{Wave_equation_axial_components_Hertzian_potential}
\left\lbrace \frac{\de ^2}{\de z^2} - \frac{1}{2}\left(\frac{f'(z)}{f(z)} \right)\frac{\de}{\de z}  + \left[\frac{\omega^2}{f(z)} - \mathbf{k}_\perp^2 - \frac{1}{2}\frac{\de}{\de z}\left(\frac{f'(z)}{f(z)} \right)\right] \right\rbrace \phi_a(z) = 0 \ ,
\end{equation}

Alternatively, for Ricci flat manifolds, using \eqref{Ricci_flat_condition}, the wave equation takes the form:

\begin{equation}\label{Wave_equation_axial_components_Hertzian_potential_Ricci_flat}
\left\lbrace \frac{\de ^2}{\de z^2} - \frac{1}{2}\left(\frac{f'(z)}{f(z)} \right)\frac{\de}{\de z}  + \left[\frac{\omega^2}{f(z)} - \mathbf{k}_\perp^2 + \frac{1}{4}\left(\frac{f'(z)}{f(z)}\right)^2\right] \right\rbrace \phi_a(z) = 0 \ .
\end{equation}

Finally, using \eqref{gauge_transformed_Faraday}, we find the Faraday 2-form:

\begin{equation}\label{Faraday_2_form_final_solution}
\begin{split}
\mathbf{F} = &\left\lbrace\left[\frac{1}{\sqrt{f(z)}}\partial_t\partial_y \right]\Pi_m - \left[\partial_x\partial_z - \frac{1}{2}\left(\frac{f'(z)}{f(z)} \right)\partial_x \right]\Pi_e \right\rbrace \de t \wedge \de x \\
- &\left\lbrace \left[\frac{1}{\sqrt{f(z)}}\partial_t\partial_x \right]\Pi_m + \left[\partial_y \partial_z - \frac{1}{2}\left(\frac{f'(z)}{f(z)}\right)\partial_y  \right]\Pi_e \right\rbrace \de t \wedge \de y \\
+ &\left\lbrace\left[\frac{1}{f(z)}\partial_t^2 - \partial_z^2 + \frac{1}{2}\left(\frac{f'(z)}{f(z)}\right)\partial_z - \frac{1}{2}\left(\frac{f'(z)}{f(z)}\right)^2 + \frac{1}{2}\left( \frac{f''(z)}{f(z)}\right) \right]\Pi_e  \right\rbrace \de t \wedge \de z \\
- &\left\lbrace\left[\frac{1}{\sqrt{f(z)}}\partial_t\partial_y \right]\Pi_e + \left[\partial_x\partial_z -  \frac{1}{2}\left(\frac{f'(z)}{f(z)} \right)\partial_x\right]\Pi_m \right\rbrace \star \de t \wedge \de x  \\
+ &\left\lbrace\left[\frac{1}{\sqrt{f(z)}}\partial_t\partial_x \right]\Pi_e - \left[\partial_y\partial_z -  \frac{1}{2}\left(\frac{f'(z)}{f(z)} \right)\partial_y\right]\Pi_m \right\rbrace \star \de t \wedge \de y \\
+ &\left\lbrace\left[\partial_x^2 + \partial_y^2 \right]\Pi_m \right\rbrace \star \de t \wedge \de z  \ .
\end{split}
\end{equation}

Using the wave equation \eqref{Wave_equation_components_Hertzian_potential_2}, we can rewrite the $\de t \wedge \de z$ component as:
  
\begin{equation}
\left\lbrace\left[\frac{1}{f(z)}\partial_t^2 - \partial_z^2 + \frac{1}{2}\left(\frac{f'(z)}{f(z)}\right)\partial_z - \frac{1}{2}\left(\frac{f'(z)}{f(z)}\right)^2 + \frac{1}{2}\left( \frac{f''(z)}{f(z)}\right) \right]\Pi_e  \right\rbrace \de t \wedge \de z = \left\lbrace\left[\partial_x^2 + \partial_y^2 \right]\Pi_e \right\rbrace \de t \wedge \de z
\end{equation}

the wave equation \eqref{Wave_equation_components_Hertzian_potential_2}, we can rewrite the $\de t \wedge \de z$ component as:
 
\begin{equation}\label{Faraday_2_form_final_solution_2}
\begin{split}
\mathbf{F} = &\left\lbrace\left[\frac{1}{\sqrt{f(z)}}\partial_t\partial_y \right]\Pi_m - \left[\partial_x\partial_z - \frac{1}{2}\left(\frac{f'(z)}{f(z)} \right)\partial_x \right]\Pi_e \right\rbrace \de t \wedge \de x \\
- &\left\lbrace \left[\frac{1}{\sqrt{f(z)}}\partial_t\partial_x \right]\Pi_m + \left[\partial_y \partial_z - \frac{1}{2}\left(\frac{f'(z)}{f(z)}\right)\partial_y  \right]\Pi_e \right\rbrace \de t \wedge \de y \\
+ &\left\lbrace \left[\partial_x^2 + \partial_y^2 \right]\Pi_e \right\rbrace \de t \wedge \de z \\
- &\left\lbrace\left[\frac{1}{\sqrt{f(z)}}\partial_t\partial_y \right]\Pi_e + \left[\partial_x\partial_z -  \frac{1}{2}\left(\frac{f'(z)}{f(z)} \right)\partial_x\right]\Pi_m \right\rbrace \star \de t \wedge \de x  \\
+ &\left\lbrace\left[\frac{1}{\sqrt{f(z)}}\partial_t\partial_x \right]\Pi_e - \left[\partial_y\partial_z -  \frac{1}{2}\left(\frac{f'(z)}{f(z)} \right)\partial_y\right]\Pi_m \right\rbrace \star \de t \wedge \de y \\
+ &\left\lbrace\left[\partial_x^2 + \partial_y^2 \right]\Pi_m \right\rbrace \star \de t \wedge \de z  \ .
\end{split}
\end{equation}

To simplify our notation, we introduce the following differential operators:

\begin{equation}\label{Diff_operators}
\begin{split}
\Da_{t i} \equiv \frac{1}{\sqrt{f(z)}}\partial_t \partial_i \quad i=x,y \ , \\
\Da_{ij} \equiv \partial_i\partial_j - \frac{1}{2}\left(\frac{f'(z)}{f(z)} \right)\partial_i  \quad i=x,y \ ,
\end{split}
\end{equation}

Using these operators, we can express the Faraday 2-form more concisely:

\begin{equation}\label{Faraday_2_form_final_solution_3}
\begin{split}
\mathbf{F} = &\left\lbrace\Da_{ty}\Pi_m - \Da_{xz}\Pi_e \right\rbrace \de t \wedge \de x 
- \left\lbrace \Da_{tx}\Pi_m + \Da_{yz}\Pi_e \right\rbrace \de t \wedge \de y 
+ \left\lbrace \nabla_s^2\Pi_e \right\rbrace \de t \wedge \de z \\
- &\left\lbrace\Da_{ty}\Pi_e + \Da_{xz}\Pi_m \right\rbrace \star \de t \wedge \de x  
+ \left\lbrace\Da_{tx}\Pi_e - \Da_{yz}\Pi_m \right\rbrace \star \de t \wedge \de y 
+ \left\lbrace\nabla_s^2\Pi_m \right\rbrace \star \de t \wedge \de z  \ .
\end{split}
\end{equation}

Interestingly, we can further simplify the wave equation \eqref{Wave_equation_components_Hertzian_potential_2}. By applying the change of variable:

\begin{equation}\label{change_of_variable}
\phi \rightarrow \sqrt{f(z)}\Pi \ ,
\end{equation}

we arrive at a new, more compact wave equation for $\Pi$:

\begin{equation}\label{new_wave_equation}
\left[\frac{1}{f}\partial_t^2 - \nabla_s^2 - \frac{1}{2}\left(\frac{f'}{f} \right)\partial_z \right] \Pi = 0 \ .
\end{equation}

This result suggests that we should reconsider our choice of the Hertzian potential, as we will explore in the next subsection.

\subsubsection{Refined Approach using Orthonormal Coframe}

To further refine our analysis, we introduce an orthonormal coframe $\left\lbrace e^0, e^1, e^2, e^3 \right\rbrace$ defined as:

\begin{equation}\label{coframe}
\begin{split}
e^0 &= \sqrt{f(l)}\de t \ , \\
e^1 &= \de x \ , \\
e^2 &= \de y \ , \\
e^3 &= \de l \ ,
\end{split}
\end{equation}

Using this coframe, we can express the metric tensor as:

\begin{equation}\label{metric_tensor_in_terms_coframe}
\mathbf{g} = - e^0 \otimes e^0 + e^1 \otimes e^1 + e^2 \otimes e^2 + e^3 \otimes e^3 \ .
\end{equation}

We now rewrite the Hertzian potential in terms of this coframe:

\begin{equation}\label{Hertzian_potential_ansatz_spatially_flat_coframe}
\begin{split}
\mathbf{\Pi} &= \Pi_e(t,x,y,l) e^0 \wedge e^3 - \Pi_m(t,x,y,l)\star e^0 \wedge e^3 \\
&= \sqrt{f(l)}\Pi_e(t,x,y,l) \de t \wedge \de l + \Pi_m(t,x,y,l)\de x \wedge \de y \ . 
\end{split}
\end{equation}

The Laplace-de Rham operator acting on the Hertzian potential becomes:

\begin{equation}\label{de_Rham__spatially_flat_coframe}
\square \mathbf{\Pi} =  \square \Pi_e  e^0 \wedge e^3 - \square\Pi_m  \star e^0 \wedge e^3  \ ,
\end{equation}

where $\square \Pi_e$ and $\square \Pi_m$ are the Laplace-de Rham operators on functions as defined in \eqref{Laplacian_functions_spatially_flat}. 

Choosing vanishing gauge terms $\mathbf{G} = \mathbf{W} = 0$, the dynamics of the Hertzian potential is given by the wave equations:

\begin{equation}\label{Wave_equation_components_Hertzian_potential_coframe}
\begin{split}
\square \Pi_e(t,x,y,l) &= \left[ \frac{1}{f(l)}\partial_t^2 - \nabla_s^2  - \frac{1}{2}\frac{f'(l)}{f(l)}\partial_l\right]\Pi_e(t,x,y,l) = 0 \ , \\
\square \Pi_m(t,x,y,l) &= \left[ \frac{1}{f(l)}\partial_t^2 - \nabla_s^2  - \frac{1}{2}\frac{f'(l)}{f(l)}\partial_l\right]\Pi_m(t,x,y,l) = 0 \ .
\end{split}
\end{equation}

Using \eqref{gauge_transformed_Faraday}, we derive the Faraday 2-form:

\begin{equation}\label{Faraday_2_form_coframe_final_solution}
\begin{split}
\mathbf{F} = &\left\lbrace \frac{1}{\sqrt{f(l)}} \partial_{ty}\Pi_m - \partial_{xl}\Pi_e \right\rbrace e^0 \wedge e^1 
- \left\lbrace \frac{1}{\sqrt{f(l)}}\partial_{tx}\Pi_m + \partial_{yl}\Pi_e \right\rbrace e^0 \wedge e^2 \\
+ &\left\lbrace \left[ \frac{1}{f(l)}\partial_t^2 - \partial_l^2 - \frac{1}{2}\frac{f'(l)}{f(l)}\partial_l \right]\Pi_e\right\rbrace e^0 \wedge e^3  \\
-& \left\lbrace\frac{1}{\sqrt{f(l)}} \partial_{ty} \Pi_e + \partial_{xl}\Pi_m\right\rbrace \star e^0 \wedge e^1  
+ \left\lbrace \frac{1}{\sqrt{f(l)}}\partial_{tx}\Pi_e - \partial_{yl}\Pi_m \right\rbrace  \star e^0 \wedge e^2 \\
+&\left\lbrace\partial_x^2 + \partial_y^2 \right\rbrace  \star e^0 \wedge e^3 \ .
\end{split}
\end{equation}

Applying the wave equation \eqref{Wave_equation_components_Hertzian_potential_coframe}, we can simplify the $e^0 \wedge e^3$ component:

\begin{equation}
\left\lbrace \left[ \frac{1}{f(l)}\partial_t^2 - \partial_l^2 - \frac{1}{2}\frac{f'(l)}{f(l)}\partial_l \right]\Pi_e\right\rbrace e^0 \wedge e^3 = \left[\left(\partial_x^2 + \partial_y^2 \right)\Pi_e \right] e^0 \wedge e^3 =  \left(\nabla_s^2 \right) e^0 \wedge e^3 \ ,
\end{equation} 

This allows us to rewrite the Faraday 2-form as:

\begin{equation}\label{Faraday_2_form_coframe_final_solution_2}
\begin{split}
\mathbf{F} = &\left( \frac{1}{\sqrt{f(l)}} \partial_{ty}\Pi_m - \partial_{xl}\Pi_e \right) e^0 \wedge e^1 
- \left(\frac{1}{\sqrt{f(l)}}\partial_{tx}\Pi_m + \partial_{yl}\Pi_e \right) e^0 \wedge e^2 \\
+ &\left[\left(\nabla_s^2 \right)\Pi_e\right] e^0 \wedge e^3  \\
-& \left(\frac{1}{\sqrt{f(l)}} \partial_{ty} \Pi_e + \partial_{xl}\Pi_m\right)\star e^0 \wedge e^1  
+ \left(\frac{1}{\sqrt{f(l)}}\partial_{tx}\Pi_e - \partial_{yl}\Pi_m \right) \star e^0 \wedge e^2 \\
+ & \left[\left(\nabla_s^2\right)\Pi_m\right] \star e^0 \wedge e^3 \ .
\end{split}
\end{equation}

To further simplify our notation, we define the following operators:

\begin{equation}\label{Diff_operators_coframe}
\begin{split}
\Da_{t i} \equiv \frac{1}{\sqrt{f(z)}}\partial_{ti} \quad i=x,y \ , \\
\Da_{ij} \equiv \partial_{ij}  \quad i=x,y \ ,
\end{split}
\end{equation}

Using these operators, we can express the Faraday 2-form more concisely:

\begin{equation}\label{Faraday_2_form_coframe_final_solution_3}
\begin{split}
\mathbf{F} = &\left\lbrace\Da_{ty}\Pi_m - \Da_{xl}\Pi_e \right\rbrace e^0 \wedge e^1 
- \left\lbrace \Da_{tx}\Pi_m + \Da_{yl}\Pi_e \right\rbrace e^0 \wedge e^2
+ \left\lbrace \nabla_s^2\Pi_e \right\rbrace e^0 \wedge e^3 \\
- &\left\lbrace\Da_{ty}\Pi_e + \Da_{xl}\Pi_m \right\rbrace \star e^0 \wedge e^1  
+ \left\lbrace\Da_{tx}\Pi_e - \Da_{yl}\Pi_m \right\rbrace \star e^0 \wedge e^2
+ \left\lbrace\nabla_s^2\Pi_m \right\rbrace \star e^0 \wedge e^3 \ .
\end{split}
\end{equation}

It is worth noting that in the limit $f(l) \rightarrow 1$, we have $\Da_{ti} \rightarrow \partial_{ti}$ and $\Da_{ij} \rightarrow \partial_{ij}$. In this case, our solution reduces to the familiar form for guided waves in flat spacetime, as given in equation \eqref{Faraday_2_form_in_terms_Hertzian_potential_flat}.

\subsubsection{TE and TM Modes}

We begin by rewriting the wave equation \eqref{Wave_equation_components_Hertzian_potential_coframe} in a more convenient form:

\begin{equation}\label{Wave_equation_components_Hertzian_potential_coframe_2}
\square \Pi(t,x,y,l) = \left[ \square_{t,l} - \nabla_\perp^2 \right]\Pi_e(t,x,y,l) = 0 \ , 
\end{equation}

where $\square_{t,l}$ is the two-dimensional Laplace-de Rham operator acting only on the temporal and axial dimensions.

As mentioned in section \ref{secc_metric}, we propose an ansatz of the form:

\begin{equation}\label{basic_ansatz_2}
\Pi(x) = A e^{-i\omega t}\psi(u,v)\phi(l) \ ,
\end{equation}

Here, $A$ is a normalization constant and $\psi(u,v)$ is an eigenstate of the transverse momentum operators $\partial_u$ and $\partial_v$. The transversal dependence of the field satisfies the scalar Helmholtz equation (as in \cite{collin1960field}):

\begin{equation}\label{Helmholtz_equation}
\left(\nabla_\perp^2 + \mathbf{k}_\perp^2\right) \psi_{\mathbf{k}_\perp^2}(u,v) = 0 \ , 
\end{equation}

where $\nabla_\perp^2 \equiv \partial_u^2 + \partial_v^2$ is the transverse Laplacian. This equation is subject to boundary conditions that constrain the set of admissible eigenvalues and eigenstates.

With this ansatz, the wave equation \eqref{Wave_equation_components_Hertzian_potential_coframe} for $\phi(l)$ becomes:

\begin{equation}\label{Wave_equation_axial_part_components_Hertzian_potential_coframe}
\left\lbrace \frac{\de^2}{\de l^2} + \frac{1}{2}\left[\frac{f'(l)}{f(l)} \right]\frac{\de}{\de l} + \left[\frac{\omega^2}{f(l)} -  \mathbf{k}_\perp^2\right] \right\rbrace \phi(l) = 0 \ ,
\end{equation}

This can be written in terms of the Laplace-de Rham operator acting on axial functions:

\begin{equation}\label{Wave_equation_deRham_axial_part_components_Hertzian_potential_coframe}
\left[\square_l - \left(\frac{\omega^2}{f(l)} -  \mathbf{k}_\perp^2\right)\right]\phi(l) = 0 \ ,
\end{equation}

or in terms of the Laplace-Beltrami operator, $\nabla^2$, acting on axial functions (minus the Laplace-de Rham operator):

\begin{equation}\label{Wave_equation_Beltrami_axial_part_components_Hertzian_potential_coframe}
- \partial_t^2 = \left[\nabla^2 + \left(\frac{\omega^2}{f(l)} -  \mathbf{k}_\perp^2\right)\right]\phi(l) = 0 \ .
\end{equation}

Alternatively, following \cite{fulling1989aspects}, we can write the wave equation \eqref{Wave_equation_components_Hertzian_potential_coframe} as:

\begin{equation}\label{Wave_equation_Fulling_form}
- \partial_t^2 \phi(x) = -f(l)\left\lbrace  \nabla_\perp^2 + \frac{\de^2}{\de l^2} + \frac{1}{2}\left[\frac{f'(l)}{f(l)} \right]\frac{\de}{\de l}  \right\rbrace \phi(x) \equiv K \phi(x) \ .
\end{equation}

With our ansatz, the operator $K$ becomes:

\begin{equation}\label{K_operator_ansatz}
K = f(l)\left\lbrace  \mathbf{k}_\perp^2 - \frac{\de^2}{\de l^2} - \frac{1}{2}\left[\frac{f'(l)}{f(l)} \right]\frac{\de}{\de l}  \right\rbrace \ .
\end{equation}

The wave equation for the axial part $\phi(l)$ can then be written:

\begin{equation}\label{Wave_equation_K_Fulling_operator_axial_part_components_Hertzian_potential_coframe}
\left(K - \omega^2 \right)\phi(l) = 0 \ .
\end{equation}

It is natural to take the eigenstates of the operator $K$ as a basis for the solutions:

\begin{equation}\label{K_operator_eigenvalue_equation}
K \phi_{\kappa_{\mathbf{k}_\perp^2}}(l) = \kappa_{\mathbf{k}_\perp^2} \phi_{\kappa_{\mathbf{k}_\perp^2}}(l) \ ,
\end{equation}

where both the eigenstates and the eigenvalues are functions of the constant transverse momentum $\mathbf{k}_\perp^2$ of the wave. The eigenvalue equation \eqref{K_operator_eigenvalue_equation} can be written as:

\begin{equation}\label{Eigenvalue_equation_axial_part}
\left\lbrace \frac{\de^2}{\de l^2} + \frac{1}{2}\left[\frac{f'(l)}{f(l)} \right]\frac{\de}{\de l} + \left[\frac{\kappa_{\mathbf{k}_\perp^2}}{f(l)} -  \mathbf{k}_\perp^2\right] \right\rbrace \phi_{\kappa_{\mathbf{k}_\perp^2}}(l) = 0 \ .
\end{equation}

An eigenstate $\phi_{\kappa_{\mathbf{k}_\perp^2}}(l)$ is a solution of the problem \eqref{Wave_equation_K_Fulling_operator_axial_part_components_Hertzian_potential_coframe} if the 'mass-shell condition' is satisfied:

\begin{equation}\label{mass_shell_condition}
\kappa_{\mathbf{k}_\perp^2} = \omega^2 \ ,
\end{equation}

In this case, both \eqref{Eigenvalue_equation_axial_part} and \eqref{Wave_equation_Beltrami_axial_part_components_Hertzian_potential_coframe} are the same equation.

Thus, a solution of the problem takes the form:

\begin{equation}
\Pi (x) = A e^{\pm i \omega t}\psi_{\mathbf{k}_\perp^2}(u,v)\phi_{\kappa_{\mathbf{k}_\perp^2}}(l) \ ,
\end{equation}

where $\psi_{\mathbf{k}_\perp^2}(u,v)$ is a solution of \eqref{Helmholtz_equation}, and $\phi_{\kappa_{\mathbf{k}_\perp^2}}(l)$ is an eigenstate of the operator K \eqref{K_operator_ansatz}, with eigenvalue $\kappa_{\mathbf{k}_\perp^2}$ satisfying the 'mass-shell condition' \eqref{mass_shell_condition}.

\paragraph{Axial Part\newline}

To analyze the axial part of the solution, we split it into a deformation of a plane wave:

\begin{equation}\label{plane_wave_deformation}
\phi_{\kappa_{\mathbf{k}_\perp^2}}(l) = q_{\kappa_{\mathbf{k}_\perp^2}}(l)e^{\pm i k_l l} \ ,
\end{equation}

This leads to the following equation for $q_{\kappa_{\mathbf{k}_\perp^2}}(l)$:

\begin{equation}\label{wave_equation_deformation_term}
\left\lbrace \frac{\de^2}{\de l^2} + \left[ \frac{1}{2}\frac{f'(l)}{f(l)} \pm 2i k_l\right]\frac{\de}{\de l}  + \left[\frac{\omega^2}{f} - \mathbf{k}^2 \pm i\frac{1}{2}\frac{f'(l)}{f(l)} k_l \right]\right\rbrace q_{\kappa_{\mathbf{k}_\perp^2}}(l) = 0 \ .
\end{equation}

Alternatively, we can express the axial part as:

\begin{equation}\label{phase_amplitude_splitting}
\phi_{\kappa_{\mathbf{k}_\perp^2}}(l) = q_{\kappa_{\mathbf{k}_\perp^2}}(l)e^{i \theta(l)} \ ,
\end{equation}

where we define:

\begin{equation}\label{phase_momentum}
k_l(l) \equiv \theta'(l) \ ,
\end{equation}

and

\begin{equation}\label{total_pseudo_momentum}
\mathbf{k}^2 = \mathbf{k}_\perp^2 + k_l^2 \ .
\end{equation}

This leads to the following equation:

\begin{equation}\label{wave_equation_phase_amplitude_splitting}
\left\lbrace \frac{\de^2}{\de l^2} + \left[ \frac{1}{2}\frac{f'(l)}{f(l)} + 2i k_l(l)\right]\frac{\de}{\de l}  + \left[\frac{\omega^2}{f} - \mathbf{k}^2(l) + i\frac{1}{2}\frac{f'(l)}{f(l)} k_l(l)  + i k_l'(l)\right]\right\rbrace q_{\kappa_{\mathbf{k}_\perp^2}}(l) = 0 \ .
\end{equation}

For propagating modes, we assume $k_l(l) \in \mathbb{R}$, and $q(l)$ to be the real amplitude of $\phi(l)$. Separating equation \eqref{wave_equation_phase_amplitude_splitting} into real and imaginary parts yields:

\begin{equation}\label{wave_equation_real_and_imaginary_parts}
\begin{split}
0 &= \frac{\omega^2}{f(l)} - \mathbf{k}_\perp^2 - k_l^2(l) +  \left(\frac{q''(l)}{q(l)} \right) + \frac{1}{2}\left(\frac{f'(l)}{f(l)} \right) \left(\frac{q'(l)}{q(l)}\right) \quad \text{(Real part)} \\
0 &=  \frac{1}{2}\left(\frac{f'(l)}{f(l)} \right)k_l(l) + k_l'(l) +  2k_l(l) \left(\frac{q'(l)}{q(l)}\right) \quad \text{(Imaginary part)} \ .
\end{split}
\end{equation}

The imaginary part provides an equation relating the phase to the amplitude:

\begin{equation}\label{phase_in_terms_amplitude}
- \frac{\de}{\de l}\left[ln (k_l) \right] = \frac{\de}{\de l}\left[ln (\sqrt{f}q^2) \right] \ ,
\end{equation}

which implies:

\begin{equation}\label{phase_in_terms_amplitude_2}
k_l(l) =  \frac{\de}{\de l} \theta(l) =   \frac{A}{\sqrt{f(l)}q^2(l)} \ ,
\end{equation}

where $A$ is an integration constant\footnote{Note that the integral expression for $\theta(l)$ appears to be a relativistic invariant}.

Substituting this into the real part of equation \eqref{wave_equation_real_and_imaginary_parts} yields:

\begin{equation}\label{amplitude_Ermakov_equation}
q''(l) + \frac{1}{2}\left(\frac{f'(l)}{f(l)} \right)q'(l) + \left[\frac{\omega^2}{f(l)} - \mathbf{k}_\perp^2\right]q(l)  = \frac{A^2}{f(l)q^3(l)}  \ ,
\end{equation}

which is a \textbf{dissipative Ermakov equation}.

\paragraph{Cuttoff Frequency\newline}

Using equation \eqref{Eigenvalue_equation_axial_part}, we can express the transversal momentum in terms of $\omega$ and $\phi$:

\begin{equation}\label{transverse_wavenumber_1}
\mathbf{k}_\perp^2 = \frac{\omega^2}{f(l)} + \frac{\phi''(l)}{\phi(l)} + \frac{1}{2}\left(\frac{f'(l)}{f(l)}\right) \frac{\phi'(l)}{\phi(l)}  \ .
\end{equation}

The momentum $\mathbf{k}_\perp^2$, also called the \textbf{cutoff wavenumber} \cite{collin1960field}, is determined by boundary conditions. The pair $\left\lbrace \omega, \mathbf{k}_\perp \right\rbrace$ determines the solution $\phi(l)$. Note that the right-hand side cannot equal $\omega^2$ as in the flat case, due to the $\frac{1}{f(l)}$ factor, which has implications for the cutoff frequency.

Alternatively, using \eqref{wave_equation_phase_amplitude_splitting}, we find:

\begin{equation}\label{transverse_wavenumber_2}
\mathbf{k}_\perp^2 = \frac{\omega^2}{f(l)} - k_l^2(l) + \frac{i}{2}\left(\frac{f'(l)}{f(l)} \right)k_l(l) + i k_l'(l) + \left(\frac{q''(l)}{q(l)} \right) + \left[\frac{1}{2}\left(\frac{f'(l)}{f(l)} \right) + 2ik_l(l) \right]\left(\frac{q'(l)}{q(l)}\right) \ .
\end{equation}

For propagating modes, $k_l(l) \in \mathbb{R}$. The wave does not propagate if $k_l(l) = 0$, defining the cutoff frequency as:

\begin{equation}\label{cutoff_wavenumber_3}
\mathbf{k}_\perp^2 = \frac{\omega^2}{f(l)}  + \left(\frac{q''(l)}{q(l)} \right) + \frac{1}{2}\left(\frac{f'(l)}{f(l)} \right)\left(\frac{q'(l)}{q(l)}\right) \ .
\end{equation}

The guide wavelength is defined as \cite{collin1960field}\footnote{This is a first-order calculation, as we seek $\Delta l$ such that $\theta \left(l + \Delta l \right) = \theta \left(l \right) + 2\pi$. Linearizing gives $\theta (l) + k_l(l) \Delta l \simeq  \theta \left(l \right) + 2\pi$, leading to $\Delta l \equiv \lambda_g \simeq  \frac{2\pi}{k_l(l)}$.}:

\begin{equation}\label{guide_wavelength}
\lambda_g \equiv \frac{2\pi}{k_l(l)}
\end{equation}

\subsubsection{TEM Modes}

TEM (Transverse Electromagnetic) modes require the transverse momentum to vanish, leading to a massless Helmholtz equation:

\begin{equation}\label{Helmholtz_equation_massless}
\nabla_\perp^2  \psi(u,v) = 0 \ .
\end{equation}

This condition allows us to describe TEM modes using either $\Pi_e$ or $\Pi_m$ \cite{rothwell2018electromagnetics}, as the vanishing transverse momentum simplifies the field structure.

The wave equation for the Hertzian potential becomes a massless Klein-Gordon equation with conformal symmetry. To exploit this, we rewrite our metric \eqref{Metric} in a conformally flat form:

\begin{equation}\label{Metric_conformal}
ds^2 =  f(l)\left[-dt^2 + \frac{1}{f(l)} dl^2\right] + du^2 + dv^2 \ .
\end{equation}

Introducing the coordinate change $dr = \frac{1}{f(l)} dl$, we obtain:

\begin{equation}
ds^2 = f(r(l))\left[-dt^2 + dr^2\right] + du^2 + dv^2 \ ,
\end{equation}

which is conformally flat in the temporal and axial dimensions. Given the conformal symmetry and the irrelevance of transverse dependence for TEM modes, we can adapt the Minkowski spacetime solutions:

\begin{equation}
\Pi(t,r) = e^{\pm i \omega t}e^{\pm i k_l r} \ .
\end{equation}

Reverting to our original coordinates yields the TEM mode solutions:

\begin{equation}\label{TEM_modes_solution}
\Pi(t,l) = e^{\pm i \omega t}e^{\pm i k_l \int^l \frac{dl'}{\sqrt{f(l')}}} \ .
\end{equation}

This solution resembles geodesic motion, as seen in radial propagation in Schwarzschild spacetime. 

\subsubsection{Application to Radial Propagation on Schwarzschild Spacetime}

We now apply our formalism to the Schwarzschild spacetime, focusing on radial propagation. Our goal is to evaluate the axial-dependent coefficient of the wave equation \eqref{Wave_equation_axial_components_Hertzian_potential}.

In Schwarzschild spacetime, the differential of proper length is given by:

\begin{equation}\label{differential_prope_length_exact_radial_Schwarzschild}
\de l = \frac{1}{\sqrt{1 - \frac{R_s}{r'}}} \de r \ .
\end{equation}

Using this, we can derive:

\begin{equation}
\frac{\de}{\de l}f(r(l)) = \partial_r f(r(l)) \frac{\de}{\de l} r(l) = \partial_r f(r(l))\sqrt{1 - \frac{R_s}{r(l)}} = \frac{r_s}{r(l)^2}\sqrt{1 - \frac{R_s}{r(l)}} \ .
\end{equation}

Hence:

\begin{equation}
\frac{f'}{f} = \frac{r_s}{r^2}\frac{1}{\sqrt{1 - \frac{r_s}{r}}} = \frac{1}{r_s}\left( \frac{r_s}{r}\right)^2\frac{1}{\sqrt{1 - \frac{r_s}{r}}} \ .
\end{equation}

Using the Taylor series \eqref{Taylor_series_1}, we can approximate this term as:

\begin{equation}
\frac{f'}{f} = \frac{1}{r_s}\left( \frac{r_s}{r}\right)^2 \left[1 + \frac{1}{2}\left(\frac{r_s}{r}\right) + \frac{3}{8}\left(\frac{r_s}{r}\right)^2 + \cdots \right] \simeq \frac{1}{r_s}\left( \frac{r_s}{r}\right)^2 \ .
\end{equation}

On the other hand:

\begin{equation}
\frac{\omega^2}{f(l)} = \omega^2 \left(1 + \left(\frac{r_s}{r}\right) + \left(\frac{r_s}{r}\right)^2 + \cdots \right)
\end{equation}

Combining these results, we can write the wave equation in the form:

\begin{equation}\label{approx_Wave_equation_axial_components_Hertzian_potential}
\left\lbrace \frac{\de ^2}{\de l^2} + \frac{1}{2r_s}\left(\frac{r_s}{r}\right)^2\frac{\de}{\de l}  + \left[\omega^2 - \mathbf{k}_\perp^2 + \omega^2\left(\frac{r_s}{r}\right) \right] \right\rbrace \phi(l) = 0 \ .
\end{equation}

Neglecting terms of order higher than $\Oa \left(\frac{r_s}{r}\right)$, we obtain:

\begin{equation}
\label{approx_Wave_equation_axial_components_Hertzian_potential_2}
\left\lbrace \frac{\de ^2}{\de l^2}  + \left[\omega^2 - \mathbf{k}_\perp^2 + \omega^2\left(\frac{r_s}{r}\right) \right] \right\rbrace \phi(l) = 0 \ .
\end{equation}

\paragraph{TEM Modes\newline}

For TEM modes, we previously derived the exact solutions \eqref{TEM_modes_solution}:

\begin{equation}\label{TEM_modes_solution_2}
\Pi(t,l) = e^{\pm i \omega t}e^{\pm i k_l \int^r(l) \de l' \frac{1}{\sqrt{f(l')}}} \ .
\end{equation}

Using \eqref{differential_prope_length_exact_radial_Schwarzschild}, the integral in \eqref{TEM_modes_solution_2} becomes:

\begin{equation}
\int^l \de l' \frac{1}{\sqrt{f(l')}} = \int^l \de r \frac{1}{f(r)} = \int^l \de r  \frac{1}{1 - \frac{R_s}{r'}} \ ,
\end{equation}

This integral has exact solution\footnote{
\begin{equation}
\int \de x \frac{1}{1 - \frac{a}{x}} = a  \ln (x-a) + x + constant 
\end{equation}
}:

\begin{equation}
\int^l \de r  \frac{1}{1 - \frac{R_s}{r'}} = R_s  \ln (l-R_s) + l + constant  \ .
\end{equation}

Consequently, the TEM mode solution \eqref{TEM_modes_solution_2} in Schwarzschild spacetime can be expressed as:

\begin{equation}\label{TEM_modes_solution_radial_Schwarzschild}
\Pi(t,l) = e^{\pm i \omega t}e^{\pm i k_l l}\left(l-R_s\right)^{\pm R_s} \ .
\end{equation}

This result reveals how the Schwarzschild geometry modifies the propagation of TEM modes, introducing a power-law correction term dependent on the Schwarzschild radius.

\subsection{Effects of Material Media}

\subsubsection{Validity of Constant Linear Constitutive Relations in Curved Spacetime}

The application of constant linear constitutive relations in curved spacetime is supported by the principle of local flatness. As noted by \cite{thompson2018covariant}, experiments probing the interaction of photons with inertia test the covariant formulation of Maxwell's equations under the assumption that local optical properties of a medium are unaffected by acceleration.
In electromagnetism, for linear, homogeneous, isotropic materials with instantaneous response to changes in electric field, we have:
\begin{equation}
\mathbf{D} = \epsilon \mathbf{E}
\end{equation}
where $\epsilon$ is a scalar permittivity. The instantaneous nature of this response suggests that it does not depend on the passage of time. Moreover, the constitutive relation is local, implying that locally we have Minkowski spacetime. Consequently, the constitutive relation should maintain the same magnitude at every point, per unit of proper length.
This motivation provides the basis for our treatment of macroscopic Maxwell equations in curved spacetime using Hertzian potentials.

\subsubsection{Solution to the Macroscopic Maxwell Equations}

Following the approach of \cite{he2016hertz}, let us consider the macroscopic Maxwell equations \eqref{Macroscopic_Maxwell_equations}:

\begin{equation}\label{Macroscopic_Maxwell_equations_2}
\begin{split}
\de \mathbf{F} &= 0 \\
\de \mathbf{G} &= \mathbf{J}_{free} \\
\mathbf{G} &= \star \mathbf{\chi} \mathbf{F} \ .
\end{split}
\end{equation}

The form of these equations suggests a general solution of the form:

\begin{equation}
\begin{split}
\mathbf{F} &= \de \delta \mathbf{\Pi}_1 - \de \mathbf{G}_1 \\
\mathbf{G} &= \de \delta \mathbf{\Pi}_2 - \de \mathbf{G}_2 \ ,
\end{split}
\end{equation}

where we have taken into account that in our case $\mathbf{J}_{free} = 0$.

While the potentials $\mathbf{\Pi}_1$ and $\mathbf{\Pi}_2$ are initially different, we can exploit the gauge freedom to set them equal. This can be achieved, for example, by writing $ \mathbf{G}_1 =  \mathbf{G}_1' + \delta \left(\mathbf{\Pi}_1 -\mathbf{\Pi}_2 \right) $. Thus, we can simplify our equations to:

\begin{equation}
\begin{split}
\mathbf{F} &= \de \delta \mathbf{\Pi} - \de \mathbf{G}_1 \\
\mathbf{G} &= \de \delta \mathbf{\Pi} - \de \mathbf{G}_2 \ . 
\end{split}
\end{equation}

Applying the constitutive equation, we can write:

\begin{equation}
\de \delta \mathbf{\Pi} - \de \mathbf{G}_2 = \star \mathbf{\chi} \left(\de \delta \mathbf{\Pi} - \de \mathbf{G}_1  \right) \ ,
\end{equation}

which implies that the potential $\mathbf{\Pi}$ satisfies the equation:

\begin{equation}
\left(1 +  \star \mathbf{\chi} \right) \de \delta \mathbf{\Pi} = \de \mathbf{G}_2 - \star \mathbf{\chi}  \de \mathbf{G}_1 \ .
\end{equation}

\subsection{Interferometry}

This section examines the interference phenomena in curved spacetime waveguides using the Hertzian potential formalism. We consider two waveguides with different proper lengths but common origin and endpoint. By superposing the solutions for individual waveguides, we derive the combined Hertzian potential and corresponding Faraday 2-form at the intersection point. This approach allows us to analyze the interference patterns that arise from the interaction of electromagnetic waves propagating through different spacetime paths.

\subsubsection{General Superposition of Guided Waves}

Consider two waveguides, labeled $a$ and $b$, with proper lengths $L_a$ and $L_b$, respectively. Both waveguides share a common origin at event $P_0$, corresponding to proper-length coordinates $l_a=l_b=0$, and extend along different paths to a common endpoint at event $P_f$, where $l_a=L_a$ and $l_b=L_b$. We describe the situation using a single coordinate time $t$ and assume that at the origin $P_0$, the coframe $\left\lbrace e^0, e^x, e^y, e^l \right\rbrace$ is identical for both waveguides.

We obtain the field equations as described previously and assume solutions of the form:

\begin{equation}\label{General_solution_two_waveguides}
\begin{split}
\mathbf{\Pi_a} &= \Pi_{E,a}(t,x_a,y_a,l_a) e^0 \wedge e^l_a - \Pi_{M,a}(t,x_a,y_a,l_a)\star e^0 \wedge e^l_a \\
\mathbf{\Pi_b} &= \Pi_{E,b}(t,x_b,y_b,l_b) e^0 \wedge e^l_b - \Pi_{M,b}(t,x_b,y_b,l_b)\star e^0 \wedge e^l_b \ ,
\end{split}
\end{equation}

where $\Pi_a$ and $\Pi_b$ represent the field states corresponding to waveguides $a$ and $b$ separately, with different metric factors $f_a(l_a)$ and $f_b(l_b)$ assumed. The components of the Hertzian potentials are the most general linear combinations of TE and TM modes:

\begin{equation}\label{Linear_combination_TE_and_TM_modes_two_waveguides}
\begin{split}
\Pi_{E,a}(t,x_a,y_a,l_a) &= \sum_{n,m} c_{E,a; n, m}\Pi_{E, a; n, m}(t,x_a,y_a,l_a)  \\
\Pi_{M,a}(t,x_a,y_a,l_a) &= \sum_{n,m} c_{M,a; n, m}\Pi_{M, a; n, m}(t,x_a,y_a,l_a) \\
\Pi_{E,b}(t,x_b,y_b,l_b) &= \sum_{n,m} c_{E,b; n, m}\Pi_{E, b; n, m}(t,x_b,y_b,l_b)  \\
\Pi_{M,b}(t,x_b,y_b,l_b) &= \sum_{n,m} c_{M,b; n, m}\Pi_{M, b; n, m}(t,x_b,y_b,l_b) \ .
\end{split}
\end{equation}

To obtain the interference pattern at $P_f$, we align the axes $e^i_a$ and $e^i_b$ at that point (performing a suitable rotation if necessary). By exploiting the linearity of Maxwell's equations (which extends to Hertzian potentials), we derive the Hertzian field at $P_f$ as a superposition:

\begin{equation}\label{Hertzian_field_superposition}
\begin{split}
\mathbf{\Pi} \left(t,x,y,L_a = L_b\right)= &\left[\sum_{n,m} \left(  c_{E,a; n, m}\Pi_{E, a; n, m} +  c_{E,b; n, m}\Pi_{E, b; n, m} \right)  \left(e^0 \wedge e^{l}\right)\right]\Big\vert_{\left(t,x,y,L_a = L_b\right)} \\ 
&- \left[\sum_{n,m} \left(  c_{M,a; n, m}\Pi_{M, a; n, m} +  c_{M,b; n, m}\Pi_{M, b; n, m} \right) \left(\star e^0 \wedge e^l_a\right)\right]\Big\vert_{\left(t,x,y,L_a = L_b\right)} \ .
\end{split}
\end{equation}

We can express this more compactly by introducing an index $w$ that sums over waveguides $a$ and $b$:

\begin{equation}\label{Hertzian_field_superposition_2}
\begin{split}
\mathbf{\Pi} \left(t,x,y,L_a = L_b\right)= &\left[\sum_{w;n,m} \left(  c_{E,w; n, m}\Pi_{E, w; n, m}\right)  \left(e^0 \wedge e^{l}\right)\right]\Big\vert_{\left(t,x,y,L_a = L_b\right)} \\ 
& - \left[\sum_{w;n,m} \left(  c_{M,w; n, m}\Pi_{M, w; n, m}  \right) \left(\star e^0 \wedge e^l_a\right)\right]\Big\vert_{\left(t,x,y,L_a = L_b\right)} \ .
\end{split}
\end{equation}

This Hertzian potential leads to the Faraday 2-form, which is also a superposition of the individual Faraday 2-forms for each waveguide:

\begin{equation}\label{Faraday_2_superposition}
\begin{split}
\mathbf{F} = & \sum_{w;n,m}\left( c_{M,w; n, m}\Da_{ty}\Pi_{M, w; n, m} -  c_{E,w; n, m}\Da_{xl}\Pi_{E, w; n, m} \right) e^0 \wedge e^1 \\ 
- &\sum_{w;n,m}\left( c_{M,w; n, m}\Da_{tx}\Pi_{M, w; n, m} +  c_{E,w; n, m}\Da_{yl}\Pi_{E, w; n, m} \right) e^0 \wedge e^2 \\
+ &\sum_{w;n,m} \left(  c_{E,w; n, m}\nabla_s^2\Pi_{E, w; n, m} \right) e^0 \wedge e^3  \\
- & \sum_{w;n,m}  \left( c_{E,w; n, m}\Da_{ty}\Pi_{E, w; n, m} + c_{M,w; n, m}\Da_{xl}\Pi_{M, w; n, m}\right) \star e^0 \wedge e^1 \\ 
+ & \sum_{w;n,m}\left( c_{E,w; n, m}\Da_{tx}\Pi_{E, w; n, m} - c_{M,w; n, m}\Da_{yl}\Pi_{M, w; n, m} \right) \star e^0 \wedge e^2 \\
+ &\sum_{w;n,m} \left(c_{M,w; n, m}\nabla_s^2\Pi_{M, w; n, m} \right) \star e^0 \wedge e^3 \ .
\end{split}
\end{equation}

This formalism for superposing Hertzian potentials from two waveguides results in a combined Faraday 2-form that describes the interference pattern at their intersection. It provides a powerful tool for studying electromagnetic interference in curved spacetime, potentially enabling the detection of subtle gravitational effects on guided waves. Future research could explore specific geometric configurations and their resulting interference patterns, offering new insights into the interplay between gravity and electromagnetism.

\section{Conclusions}\label{conclusions_section}
In this work we have developed a framework for analyzing guided electromagnetic waves in static curved spacetimes using Hertzian potentials. Our key findings include:
\begin{itemize}
    \item The development of a formalism for studying interference patterns in curved spacetime waveguides, potentially enabling new precision tests of general relativity.
    \item The derivation of wave equations for TE and TM modes in curved spacetime, generalizing the flat spacetime results.
\end{itemize}

The analysis of guided waves in static curved spacetimes using Hertzian potentials has yielded several key results:
\begin{itemize}
\item  The wave equation for the axial part of the Hertzian potential in curved spacetime (equation \eqref{Wave_equation_axial_part_components_Hertzian_potential_coframe}) generalizes the flat spacetime case, adding a dependance on the metric function $f(l)$.
\item  The transverse part of the field satisfies a scalar Helmholtz equation (equation \eqref{Helmholtz_equation}), similar to the flat spacetime case.
\item  The axial part of the solution can be expressed as a deformation of a plane wave (equation \eqref{plane_wave_deformation}), leading to a dissipative Ermakov equation for the amplitude (equation \eqref{amplitude_Ermakov_equation}):
   \begin{equation}
   q''(l) + \frac{1}{2}\left(\frac{f'(l)}{f(l)} \right)q'(l) + \left[\frac{\omega^2}{f(l)} - \mathbf{k}_\perp^2\right]q(l)  = \frac{A^2}{f(l)q^3(l)}  \ .
   \end{equation}
\item  The cutoff wavenumber in curved spacetime (equation \eqref{cutoff_wavenumber_3}) depends on the metric function and its derivatives, generalizing the flat spacetime result.
\item  For TEM modes, the solution (equation \eqref{TEM_modes_solution}) is a simple modification of the flat spacetime case, with the phase corrected by the proper time along the waveguide:
   \begin{equation}
   \Pi(t,l) = e^{\pm i \omega t}e^{\pm i k_l \int^l \frac{dl'}{\sqrt{f(l')}}} \ .
   \end{equation}
\item In the specific case of radial propagation in Schwarzschild spacetime, the TEM mode solution (equation \eqref{TEM_modes_solution_radial_Schwarzschild}) includes a power-law correction term dependent on the Schwarzschild radius:
   \begin{equation}
   \Pi(t,l) = e^{\pm i \omega t}e^{\pm i k_l l}\left(l-R_s\right)^{\pm R_s} \ .
   \end{equation}
\end{itemize}

These results demonstrate how spacetime curvature modifies the propagation of electromagnetic waves in waveguides, providing a foundation for studying guided waves in more general curved spacetime scenarios.

Future work could explore specific applications of this framework, such as precision tests of gravitational effects on electromagnetic waves in long-baseline experiments. Additionally, extending this analysis to dynamic spacetimes and investigating the interplay between gravity and quantum optics in waveguides present exciting avenues for further research.

As future research directions and TODO items, we propose:

\begin{itemize}
    \item Complete the theory for propagation in material media, extending our analysis to include the effects of dielectric and magnetic materials in curved spacetime.
    \item Derive applications with observables that are dependent on curved spacetime, beyond the proper time differences explored in previous work \cite{Zych_interferometry,Zych_thesis}. In particular, investigate the effects on TE and TM modes, which could provide novel ways to detect spacetime curvature.
    \item Explore observables on quantum states (photons) that are influenced by curved spacetime. This could involve adapting classical setups like the Michelson-Morley experiment to the quantum regime in curved spacetime waveguides. For example, building upon the work of Izmailov et al. \cite{izmailov1993stationary} and Kowalski \cite{kowalski1992gravitational}, one could investigate the use of fiber optics or dielectrics in a Michelson interferometer to enhance the gravitational effects on light propagation.
\end{itemize}

These proposed extensions of our work have the potential to bridge the gap between classical and quantum effects in curved spacetime, possibly leading to new experimental tests of general relativity and quantum mechanics in curved spacetime.

\pagebreak

\section{Code Availability}

Companion Maple Worksheets with some of the symbolic calculations in Section \ref{guided_waves_static_st_section} can be found in the GitHub repository \cite{BlancoRomero2024}.

\section{Acknowledgements}

This work was conducted during a research internship in the Department of Optics of Universidad Complutense de Madrid between 2017 and 2018, in the context of the \textit{Quantum Metrology for General Relativity} project (Spanish Ministerio de Educación, Cultura y Deporte (MECD) \textit{Collaboration Grant} for the 2017-2018 academic year). I would like to thank Professor Alfredo Luis Aina for his valuable support and insightful discussions throughout the internship. I would also like to thank Daniel Sobral-Blanco for his fruitful comments.

\printbibliography

\end{document}